\documentclass[twocolumn]{aastex631}
\usepackage{threeparttable}
\usepackage{parskip}
\usepackage{float}
\usepackage{hyperref}
\usepackage{graphicx}

\usepackage{booktabs}

\usepackage{amsthm}
\usepackage{amsmath}

\usepackage{soul}

\begin{document}
%\linenumbers
\title{Broadband Variability Analysis of FSRQ PKS\,0402-362  with Indications of Quasi-Periodic Modulation}

\author{Zeeshan Nazir}
\affiliation{Department of Physics, Central University of Kashmir, Ganderbal 191201, India.}
\email{mirzeeshannazir@gmail.com}
\author{Sikandar Akbar}
\affiliation{Department of Physics, University of Kashmir, Srinagar 190006, India.}
\email{darprince46@gmail.com}
\author{Zahir Shah}
\affiliation{Department of Physics, Central University of Kashmir, Ganderbal 191201, India.}
\email{shahzahir4@gmail.com}
\author{Athar A. Dar}
\affiliation{Department of Physics, Central University of Kashmir, Ganderbal 191201, India.}
\affiliation{Department of Physics, University of Kashmir, Srinagar 190006, India.}
\email{ather.dar6@gmail.com}
\author{Zahoor Malik}
\affiliation{Department of Physics, National Institute of Technology, Srinagar~190006, India.}
\email{malikzahoor313@gmail.com}

\begin{abstract}
We present a comprehensive temporal and spectral study of the flat-spectrum radio quasar PKS~0402$-$362 using  \textit{Fermi}-LAT/Swift-XRT/UVOT observations spanning from MJD 54686-60321. The $\gamma$-ray light curve exhibits multiple phases of enhanced activity, with the fractional variability parameter ($F_{\mathrm{var}}$) showing larger amplitudes at longer timescales, consistent with variability trends observed in other FSRQs. Statistical analysis of the flux and spectral index distributions using the Anderson--Darling test and histogram fitting reveals that both distributions deviate from a single log-normal form and are better represented by a double log-normal profile, indicating two distinct flux states.
A search for quasi-periodic oscillations  in the $\gamma$-ray emission using the Lomb--Scargle periodogram identified a significant periodic signal at $\sim$413~days with a confidence level exceeding $3\sigma$. However  the proximity of the timescale to one
year and  limited number of observed cycles prevents a definitive interpretation.
Broadband spectral energy distributions for six flux states were modeled using a one-zone leptonic framework incorporating synchrotron, synchrotron self-Compton (SSC), and external Compton (EC) components. The SEDs are well reproduced with physically reasonable parameters: high-flux states exhibit harder electron spectra and lower magnetic field strengths ($B \sim 0.2$--$0.6$\,G), while low-flux states show softer spectra and stronger magnetic fields ($B \sim 1.3$\,G). The fitted break energy decreases during high-flux states, suggesting enhanced radiative cooling and a transition toward a particle- or kinetic-energy dominated jet. These trends are consistent with the “harder-when-brighter” behavior commonly observed in blazars.
\end{abstract}


\section{Introduction}
The field of $\gamma$-ray astronomy has seen remarkable growth over the past 1.5 decades, largely due to the capabilities of the Fermi Large Area Telescope (LAT) instrument. The latest Fermi catalogue release (4FGL DR3) includes over 5000 gamma-ray sources, with a striking 90 per cent identified as blazars. Blazars are a unique type of active galactic nucleus (AGNs) distinguished by a powerful relativistic jet of plasma directed towards the observer's line of sight \citep{1979}. Their emission spans across the electromagnetic spectrum from radio to $\gamma$-ray energies, making them some of the brightest sources in the $\gamma$-ray universe. Besides their broad emission spectrum, blazars exhibit significant variability, with flux doubling times ranging from minutes to days \citep{2007}. This intense behavior is generally attributed to the relativistic movement of the emission region along the jet, which helps in constraining the source’s energy dynamics. %Based on the presence or absence of line features in their optical spectrum, blazars are further classified into flat spectrum radio quasars (FSRQs) and BL Lac objects.\\
Blazar class is further categorized into two sub-classes: Flat Spectrum Radio Quasars (FSRQs) and BL Lacertae objects (BL Lacs) according to the presence (in the former) or absence (in the latter) of emission lines in their optical spectrum \citep{1995PASP..107..803U}.\\

The spectral energy distribution (SED) of blazars typically features two prominent peaks. The low-energy peak is well understood to arise from synchrotron radiation emitted by a non-thermal electron population. In contrast, the high-energy peak is generally attributed to synchrotron self-Compton (SSC) processes and/or external Compton (EC) scattering, where the same electrons scatter external photons to high energies \citep{1985}. For EC processes, these photons can originate from various sources, such as the broad-line regions, thermal infrared radiation from a dusty torus, or even the accretion disc \citep{1993} .
The broad-band SED of BL Lac objects is well explained by synchrotron and SSC processes alone \citep{2008}. In contrast, for FSRQs, simultaneous X-ray and $\gamma$-ray observations indicate that both SSC and EC processes are necessary to account for their high-energy emissions \citep{sahayanathan-2012, 10.1093/mnras/stx1194,  10.1093/mnras/stad3818,Akbar_2024}. Beyond these leptonic emission models, the high-energy component in blazars has also been also interpreted as an outcome of hadronic cascades \citep{1992}.\\
%The field of $\gamma$-ray astronomy has seen remarkable growth over the past 1.5 decades, largely due to the capabilities of the Fermi-LAT instrument. The latest Fermi catalogue release (4FGL DR3) includes over 5000 gamma-ray sources, with a striking 90 per cent identified as blazars.
 %This vast dataset offers a unique foundation for exploring fundamental questions in high-energy astrophysics, such as the dominant mechanisms driving gamma-ray production, the processes that accelerate particles to extreme energies, the localization of emission sites 
%within jets, and the interplay between emissions at different wavelengths. 

Blazar variability---particularly in the $\gamma$-ray regime---has been found to exhibit a lognormal distribution, 
indicating the presence of multiplicative processes such as turbulence and particle cascades within relativistic jets 
\citep{Shah_2018, 2020MNRAS.496.3348S}. This lognormal trend implies that the observed flux variations in blazars are not 
purely stochastic but arise from dynamic physical processes occurring within the jet. Energy dissipation through 
turbulence, driven by chaotic motions in the jet, and particle cascades involving interactions between energetic 
particles and external photon fields, play significant roles in producing the observed $\gamma$-ray variability 
\citep{2010MNRAS.402..497G}. Moreover, variations in the photon index---which characterizes the spectral slope---are often 
correlated with changes in flux, highlighting the strong connection between particle acceleration mechanisms and the 
energy distribution within the jet \citep{2010ApJ...716...30A}.

The search for quasi-periodic oscillations (QPOs) in blazars provides a powerful diagnostic tool for probing the 
physical processes operating in the vicinity of supermassive black holes (SMBHs), including jet precession, 
accretion disk instabilities, and interactions within binary black hole systems \citep{zxgv-fzv5,sharma2023detection}.

The flat-spectrum radio quasar PKS~0402$-$362 (catalogued as 4FGL~J0403.9$-$3605, $z = 1.417$) has exhibited remarkable variability across multiple epochs in the $\gamma$-ray regime. The first significant detection was reported by the Fermi-LAT 
and the AGILE satellite in February~2010, when enhanced $\gamma$-ray activity was observed with a high-state flux 
above 100~MeV measured at approximately $(1.0 \pm 0.3) \times 10^{-6}$~ph~cm$^{-2}$~s$^{-1}$ 
\citep{2010ATel.2413....1H}. Shortly thereafter, during 13-16~March~2010, AGILE confirmed continued activity 
from the source, reporting an increasing flux with a detection significance of about $5\sigma$ 
\citep{2010ATel.2484....1S}. A major outburst was subsequently detected by Fermi-LAT in September~2011 
\citep{2011ATel.3655....1D}, followed by Swift Target of Opportunity observations that revealed corresponding 
X-ray emission, suggesting a correlated multiwavelength flare event. Near-infrared monitoring under the TANAMI program 
using the REM telescope also recorded a flux enhancement nearly coincident with the 2011 $\gamma$-ray flare, reinforcing 
the contemporaneous origin of the emissions \citep{2011ATel.3660....1N}. Another episode of $\gamma$-ray brightening 
was observed in August~2014, which prompted further Swift follow-up observations to investigate its spectral and 
temporal behavior \citep{2014ATel.6391....1O}. More recently, both AGILE and Fermi-LAT 
 reported renewed high-energy activity from PKS~0402$-$362 in February~2023, confirming its persistent 
variability and long-term energetic nature 
\citep{2023ATel15903....1C,2023ATel15905....1B,2023ATel15952....1P}. 
\citet{2023MNRAS.521.3451D} presented a detailed study of PKS~0402$-$362 using twelve years of \textit{Fermi}-LAT observations, covering August~2008 to January~2021. They identified three major epochs of elevated $\gamma$-ray flux, with the second phase representing the brightest ever recorded from this source. %The asymmetric flare profiles were interpreted as signatures of slow particle cooling or variations in the Doppler boosting factor as the main drivers of the observed variability. Their spectral fitting with power-law and log-parabola models revealed steep spectral indices, while broadband SED modeling supported a single-zone leptonic emission scenario with particle acceleration consistent with diffusive shock acceleration processes.

In this study, we present a comprehensive investigation of PKS~0402$-$362, 
focusing on its long-term variability and emission characteristics across multiple wavelengths. The analysis 
utilizes all publicly available data from the Fermi-LAT and the Swift 
observatory, spanning the period from August~2010 to March~2024, thereby encompassing multiple epochs of both 
high and low activity states of the source. Since its inclusion in the Fermi-LAT all-sky monitoring 
program in August~2008, PKS~0402$-$362 has exhibited several episodes of intense $\gamma$-ray activity, making 
it an excellent target for probing the physical processes governing blazar variability. 
The temporal variability of PKS~0402$-$362 is quantified through fractional variability analysis, providing a 
measure of the amplitude of flux variations across different energy bands and epochs. To investigate the 
statistical nature of the emission, we examine the distribution of $\gamma$-ray fluxes, testing for lognormality 
and potential non-linear variability patterns that are often indicative of stochastic processes within blazar jets. 
We also  performed a systematic search for QPOs on the $\gamma$-ray light curve 
to identify possible signatures of periodic or quasi-periodic behavior that may arise from jet precession, 
accretion-disk instabilities, or binary supermassive black hole dynamics. 
We also investigate the properties of the highest-energy photon associated with PKS~0402$-$362 detected during 
the LAT monitoring period. The energy and arrival time of this photon provide important constraints on the location 
of the emission region within the relativistic jet and on the underlying particle acceleration mechanisms 
responsible for producing very high-energy $\gamma$-rays.

In addition to variability and temporal analyses, we also perform SED modeling of PKS~0402$-$362 during distinct flux states to understand the underlying emission mechanisms. The SEDs are constructed using simultaneous $\gamma$-ray, X-ray, and optical/UV observations obtained from Fermi-LAT and Swift-XRT/UVOT. Each flux interval is modeled within a one-zone leptonic framework, where synchrotron, SSC, and EC processes jointly reproduce the observed spectra. The resulting fits constrain key jet parameters such as the magnetic field strength, bulk Lorentz factor, and electron energy distribution indices, providing insights into particle acceleration and radiative processes within the jet. This comprehensive approach links temporal variability with spectral evolution, enabling a unified understanding of the physical conditions driving the multi-wavelength emission in PKS~0402$-$362.

\section{Observations and Data Analysis}\label{sec:obs_data} \label{sec:observation_data_analysis}

\subsection{Fermi-LAT}
Fermi large area telescope (Fermi-LAT) is a pair conversion detector \citep{Atwood_2009} with large effective area ($\sim8000 \,\text{cm}^2/GeV $photon) and large field of view ($\sim$2.4 sr). LAT is sensitive to photons with energy ranging from 20 MeV to 500 GeV\label{sec:maths}. % used for referring to this section from elsewhere
Fermi-LAT has been designed to observe $\gamma$-rays across the energy range of 20 MeV to over 300 GeV \citep{Atwood_2009}. 
%It represents a collaborative effort involving NASA, the U.S. Department of Energy (DOE), and various research institutions from France, Italy, Japan, and Sweden.
Operating in its standard scanning mode, Fermi surveys the entire sky every 3 hours.

In this study, we utilized $\gamma$-ray data for PKS 0402-362 collected by Fermi-LAT during the period from MJD --- (August 2010 to March 2024). We utilized weekly binned $\gamma$-ray light curves available from the Fermi-LAT Light Curve Repository \citep{Baldini_2021} for the temporal analysis.
For the spectral analysis, the data was processed using FERMITOOLS-v2.0.1 following the standard analysis procedures outlined in the Fermi-LAT documentation.  We used the latest Fermi-LAT 4FGL catalogue to obtain the XML model file, which includes all sources within the region of interest (ROI) along with their spectral
models, positions, and normalizations. Specifically, P8R3 events were extracted from a 15° region of interest centered on the source. Events likely to be photons were selected using the SOURCE class parameters (evclass=128 and evtype=3), while photons arriving from zenith angles greater than 90° were excluded to avoid contamination from Earth limb $\gamma$-rays. The Galactic diffuse emission was modelled using gll$_-$iem$_-$v07.fits, and the isotropic emission component was accounted for with iso$_-$P8R3$_-$CLEAN$_-$V3$_-$v1.txt. Furthermore, the analysis used the post-launch instrument response function P8R3$_-$SOURCE$_-$V3. For the SED modeling, the flux points
and energies from Fermi-LAT observations were converted into XSPEC-readable (PHA) format using the ftflx2xsp tool.
\subsection{Swift-XRT} 
The X-ray observations used in this study were obtained with the X-Ray Telescope (XRT; \citealt{2004ApJ...611.1005G}) onboard the Neil Gehrels Swift Observatory. Swift carried out a total of 30 observations of PKS~0402$-$362
 between MJD 55218 and 60025. Each observation corresponds to a single data point in the X-ray light curve, establishing a direct mapping between the observation IDs and the temporal evolution of the source.
The data, collected in photon-counting (PC) mode, were processed using the XRTDAS software package (version 3.0.0), integrated within the HEASOFT suite (version 6.27.2). Following the standard Swift analysis procedure, the XRTPIPELINE task (version 0.13.5) was used to generate Level 2 event files, ensuring cleaned and calibrated data. Source spectra were extracted from a circular region of radius 50 arcsec centered on the source, while background spectra were obtained from a nearby, source-free circular region with a radius of 100 arcsec.
Ancillary response files (ARFs) were generated using the XRTMKARF tool, and the spectra were grouped using GRPPHA to ensure a minimum of 20 counts per bin for valid $\chi^2$ statistics. Spectral fitting was performed in XSPEC (version 12.11.0; \citealt{xspec}) using an absorbed power-law model. Galactic absorption was accounted for using the Tbabs model, which incorporates the effects of neutral hydrogen absorption. The hydrogen column density ($n_{\mathrm{H}}$) was fixed at $6.92 \times 10^{19}\,\mathrm{cm^{-2}}$, as reported by \citet{2005A&A...440..775K}.

\subsection{Swift-UVOT}
The Ultraviolet/Optical Telescope (UVOT; \citealt{2005SSRv..120...95R}) onboard the Neil Gehrels Swift Observatory provides optical and ultraviolet (UV) observations through six filters: v, b, u, w1, m2, and w2 \citep{2008MNRAS.383..627P}. To process the UVOT data for PKS~0402$-$362, we used the HEASOFT software package (version 6.26.1). The available frames for each filter were summed using the UVOTIMSUM task to obtain a single combined image per filter. In cases where only a single frame was available, that frame was used directly for analysis.
Image processing and photometric extraction were performed using the UVOTSOURCE task. Source counts were extracted from a circular region with a radius of 5 arcsec, while the background was estimated from a nearby, source-free circular region of radius 10 arcsec.
To correct for Galactic extinction, we adopted the values from \citet{2011ApJ...737..103S}, applying an extinction correction based on $E(B-V) = 0.0044$ and using $R_V = A_V / E(B-V) = 3.1$.
\begin{figure*}
\centering
\includegraphics[width=1.0\linewidth]{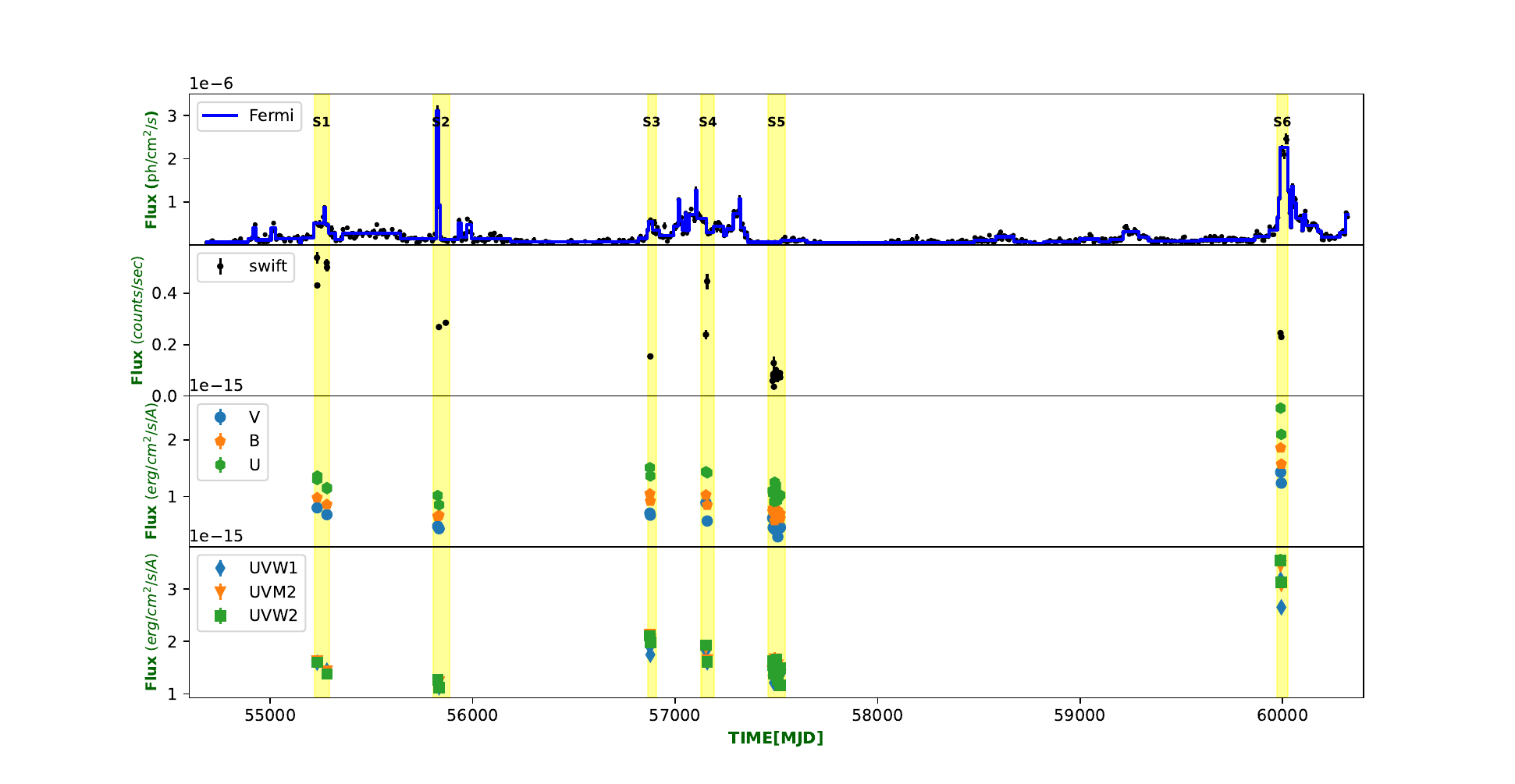}
\caption{Multiwavelength light curve of FSRQ PKS~0402$-$362 in different flux states. The top panel of the multiplot presents the 7-day binned $\gamma$-ray light curve, integrated over the 0.1–100 GeV energy range. The subsequent panels—second, third, and fourth—show the X-ray, UV, and optical light curves, respectively. The colored vertical bands mark the intervals where broadband spectral modeling has been carried out.
}
\label{fig:multiplot}
\end{figure*}
\section{Temporal Analysis}\label{sec:temp}

To investigate the temporal characteristics of PKS~0402$-$362, we generated its multiwavelength light curve in the UV/optical, X-ray, and $\gamma$-ray energy bands covering the period MJD~54686-60321. All available observations during this interval were utilized, and the corresponding $\gamma$-ray, X-ray and UV/optical datasets were processed following the methods described in Section~\ref{sec:obs_data}. The resulting multiwavelength light curve is shown in Figure~\ref{fig:multiplot}, where the top panel presents the weekly binned $\gamma$-ray (0.1--100~GeV), and the second, third, and fourth panels depict the X-ray, UV, and optical light curves. Each point in the X-ray and optical/UV curves represents a single observational epoch.

\begin{figure}
\centering
\includegraphics[width=1.0\linewidth]{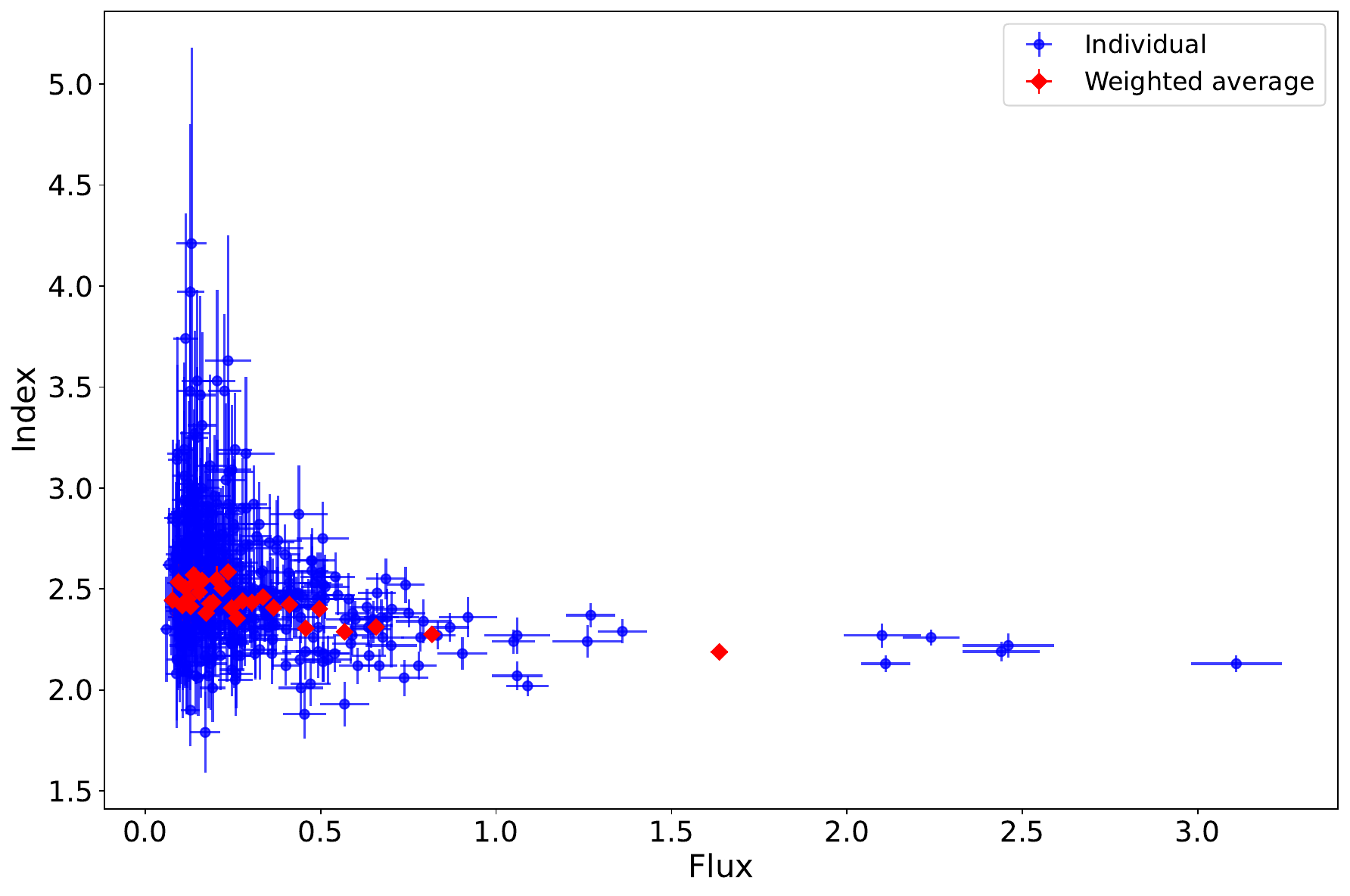}
\caption{Correlation plot between index and flux in 7-day binned light curve. The red diamond symbols denote the weighted mean index values plotted against the corresponding weighted mean flux values.}
\label{fig:flux-index}
\end{figure}

% The weekly binned $\gamma$-ray light curve is obtained by integrating photons within 0.1-300 GeV. The multi-wavelength light curve, constructed using data from Fermi-LAT and Swift-XRT/UVOT observations, is presented in 
 
The source was in an elevated $\gamma$-ray emission state on 2023 February 13, with a maximum weekly  $\gamma$-ray flux (E>100 MeV) of $(3.11 \pm 0.13) \times 10^{-6} \, \mathrm{ph \, cm^{-2} \, s^{-1}}$ at MJD 55827.5. This corresponds to a flux increase of factor 20 relative to the average flux reported in the fourth Fermi-LAT catalog (4GFL-DR3) \citep{Abdollahi_2022}. 
The correlation between the weekly binned $\gamma$-ray flux and the spectral index, illustrated in Figure~\ref{fig:flux-index}, indicates a  ``harder-when-brighter" trend in the source. To further examine this relation, the flux and index values were arranged in ascending order of flux and subsequently averaged over bins of 13 data points. In Figure~\ref{fig:flux-index}, the red diamond symbols denote the weighted mean index values plotted against the corresponding weighted mean flux values. Applying the Spearman rank correlation test yielded a coefficient of $-0.63$ and a null-hypothesis probability of $2.15 \times 10^{-4}$. These results indicate a moderate ``harder-when-brighter" behaviour in PKS\,0402$-$362, a characteristic frequently observed in blazars \citep{10.1093/mnras/stz151, Britto_2016}.  Since the Fermi-LAT energy range is situated near the peak of the inverse IC component in the SED, spectral hardening during flaring states suggests an increase in the detection of higher-energy photons. This behaviour could occur if the IC peak of the SED shifts to higher energies during periods of elevated flux. Notably, a similar feature with a strong correlation between spectral hardening during flares and shifts in the IC peak energy was observed by \citet{10.1093/mnras/stz151}.

The multiwavelength plot (see Figure \ref{fig:multiplot}) exhibits fluctuations in flux across various energy ranges. %Blazars exhibit variability across all observed energy bands, with these variations becoming particularly pronounced during flaring events. 
The amplitude of variability is influenced by key source parameters such as the magnetic field, viewing angle, and particle density \citep{refId0}. The availability of multi-wavelength data in the $\gamma$-ray, X-ray, and UV/optical bands for PKS\,0402$-$362 enables a comparative analysis of variability amplitudes across different energy ranges. The fractional variability amplitude was computed using the method described by \citep{10.1046/j.1365-2966.2003.07042.x}.\\
\begin{figure}
\centering
\includegraphics[width=0.9\linewidth]{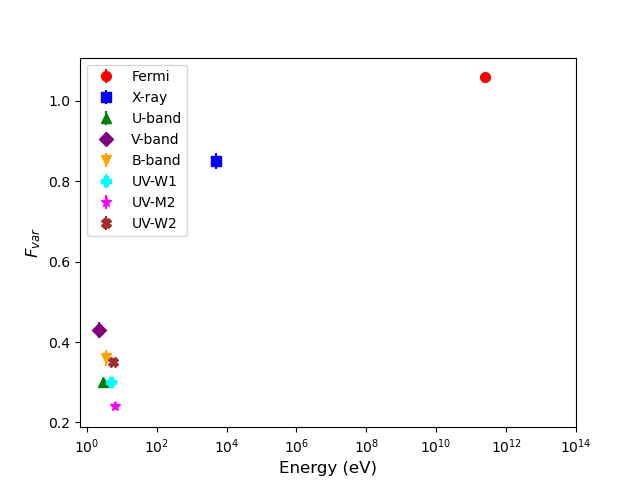}
\caption{Energy-dependent fractional variability in different energy bands}.
\label{fig:energy_var}
\end{figure}

\begin{eqnarray}
F_{\rm var} = \sqrt{\frac{S^2 - \langle \sigma_\mathrm{err}^2 \rangle}{\langle F \rangle^2}} 
\end{eqnarray}
where $S^2$, $\langle F \rangle$, $\langle \sigma_\mathrm{err}^2 \rangle$, and are the flux variance, mean flux, and mean square error in the measurement of flux, respectively. The uncertainty in $F_{\rm var}$ is obtained by the equation
%\begin{eqnarray}
%\Delta F_\mathrm{var} = \sqrt{F_\mathrm{var}^2 + {\rm err}(\sigma_\mathrm{NXS}^2)} - F_\mathrm{var} ,
%\end{eqnarray}
%where,
\begin{eqnarray}
\Delta F_\mathrm{var} = \sqrt{ \left(\sqrt{\frac{1}{2N}} \cdot \frac{\langle \sigma _\mathrm{err}^{2}\rangle }{F_{var}\langle F \rangle ^{2}} \right)^{2} + \frac{\langle \sigma _\mathrm{err}^{2}\rangle }{N\langle F \rangle ^{2}}} ,
\end{eqnarray}

Here, N represents the total number of flux points in the light curve. 
\begin{table}
    \centering
    \begin{tabular}{c|c}
      Energy band (eV)   &  Variability Amplitude \\ \hline
      $\gamma$-ray (0.1-100 GeV)  &   $1.06\pm 0.01$ \\
      X-ray (0.3--10~keV) & $0.85 \pm 0.02$ \\

      U  &   $0.30\pm0.01$ \\
      V &     $0.43\pm0.02$ \\
      B &     $0.36\pm0.02$  \\
      UVW1  &  $0.30\pm0.01$  \\
      VVM2 &    $0.24\pm0.01$  \\
      UVW2 &    $0.35\pm0.01$ 
    \end{tabular}
    \caption{Fractional varaibility amplitude (Fvar) of source in different energy bands with simutaneous data across the light curve}
    \label{tab:frac_var}
\end{table}

The computed values of $F_{\mathrm{var}}$, together with their corresponding uncertainties for the $\gamma$-ray, X-ray, and UV/optical bands, are listed in Table~\ref{tab:frac_var} and illustrated in Figure~\ref{fig:energy_var}. The lowest value of $F_{\mathrm{var}}$ is observed in the optical/UV band, while the highest corresponds to the $\gamma$-ray energy range. This behaviour is contrary to the general trend observed in FSRQs, where the X-ray band typically exhibits the lowest variability. We note, however, that the present analysis is based on sparsely sampled and non-continuous data, particularly in the X-ray and optical/UV bands, and hence the variability trend reported here should be interpreted with caution. If the observed variability is associated with changes in the jet parameters, such as the bulk Lorentz factor, one would expect comparable $F_{\mathrm{var}}$ values in the optical/UV and X-ray bands, assuming that the emission in these energies arises from synchrotron and SSC processes \citep{2018RAA....18...35S}. The optical/UV emission in FSRQs often includes a substantial thermal contribution from the accretion disk, which remains relatively stable on short timescales compared to the non-thermal jet emission. This can dilute the jet-related variability, resulting in a smaller observed $F_{\mathrm{var}}$ in these bands. In contrast, the $\gamma$-ray emission originates from high-energy electrons, whose rapid radiative cooling and variations in the acceleration processes can lead to stronger flux variability.

%Such behaviour has also been reported for several other blazars \citep{Balokovic_2016,ccc,10.1093/mnras/stw2342,10.1093/mnras/stab834,10.1093/mnras/stac1616}. 
%If the observed variability is associated with changes in the jet parameters, such as the bulk Lorentz factor, one would expect comparable $F_{\mathrm{var}}$ values in the optical/UV and X-ray bands (assuming that the emission in these energies arises from synchrotron and SSC processes \citep{2018RAA....18...35S}). On the other hand, the large difference in $F_{\mathrm{var}}$ between the optical/UV and $\gamma$-ray bands relative to the X-ray band can be interpreted in terms of changes in the emitting electron distribution. The X-ray emission may correspond to the low-energy tail of the Compton component, whereas the optical/UV and $\gamma$-ray emissions lie at the high-energy ends of the synchrotron and Compton components (see Section~\ref{sec:sed}). Consequently, the X-ray photons originate from relatively low-energy electrons compared to those producing the optical/UV and $\gamma$-ray emissions. Since the radiative loss rate scales with the square of the electron energy, high-energy electrons lose energy more rapidly than low-energy ones, leading to enhanced variability in the $\gamma$-ray and optical/UV bands compared to X-rays.

The $F_{\mathrm{var}}$ values obtained for the PKS\,0402$-$362 (see Table \ref{tab:fvar_cut}) exhibit a consistent dependence on both time binning and data selection. When all flux points are included, $F_{\mathrm{var}}$ increases from the 3-day to the 7-day bin and then shows a slight decrease at the 30-day bin. After applying the flux significance cut ($\mathrm{flux/flux_{error} > 3}$), which retains only statistically reliable data and minimizes bias from low signal-to-noise measurements, the variability amplitude displays a steady increase from the 3-day to the 30-day bin. This trend indicates that the variability of FSRQs becomes more pronounced at longer timescales. Such behavior is consistent with the ensemble results reported by \citet{2025PhRvD.111l3052S}, where FSRQs showed a gradual rise in $F_{\mathrm{var}}$ with increasing bin size, and with the findings of \citet{Akbar_2024}, who demonstrated that the FSRQ source PKS\,0805-07 exhibits a similar increase in $F_{\mathrm{var}}$ when considering data points with $TS > 4$. In contrast, \citet{2019Galax...7...62S} reported an opposite trend for BL~Lac sources, where $F_{\mathrm{var}}$ decreases with larger time bins when all data points are included, underscoring the strong sensitivity of $F_{\mathrm{var}}$ to data quality and source class.

\begin{table}
\centering
\caption{Fractional variability ($F_{\mathrm{var}}$) with and without flux cut for different time bins.}
\label{tab:fvar_cut}
\begin{tabular}{lcccc}
\hline
Time Bin & \multicolumn{2}{c}{Without Cut} & \multicolumn{2}{c}{With Cut (flux/flux\_error > 3)} \\

\cline{2-3} \cline{4-5}
 & $F_{\mathrm{var}}$ & $F_{\mathrm{var,err}}$ & $F_{\mathrm{var}}$ & $F_{\mathrm{var,err}}$ \\
\hline
3-day  & 1.24 & 0.01 & 0.94 & 0.01 \\
7-day  & 1.31 & 0.01 & 1.06 & 0.01 \\
30-day & 1.22 & 0.01 &  1.13 & 0.01 \\
\hline
\end{tabular}
\end{table}

\subsection{Flux Distribution}\label{sec:dist}
The examination of flux and index distributions in astrophysical systems serves as a valuable tool for probing the underlying physical processes responsible for their variability. To investigate the statistical behaviour of the flux and spectral index values, we employed the Anderson–Darling (AD) test along with histogram fitting techniques. The AD test yielded test statistic (TS) values of 8.6 and 4.9 for the flux and index distributions, respectively. Since these TS values exceed the critical value at the 5\% significance level, it implies that neither the flux nor the index distributions follow a normal or log-normal profile. Consequently, we further examined the possibility of these distributions exhibiting bimodal characteristics.

\begin{figure}
\centering
\includegraphics[width=1.1\linewidth]{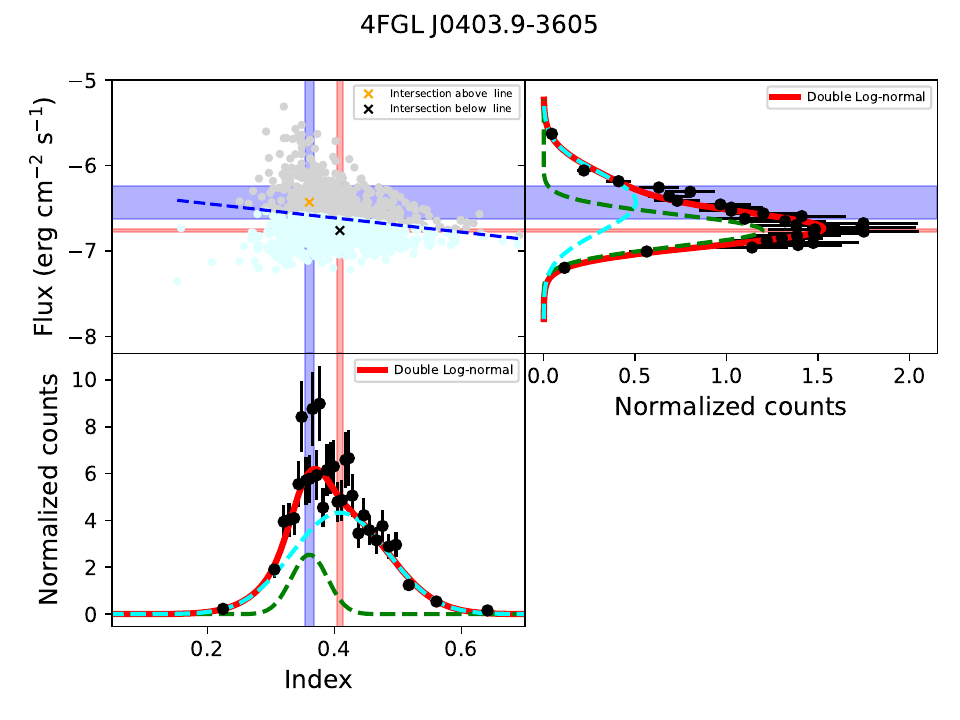}
\caption{Multiplot showing the flux/index distribution of PKS\,0402-362 in 7-day binning. The top panel shows the scatter plot of the logarithm of flux verses index, along with the best fit line(dotted blue).Grey and light cyan points are above and below the best fit line, respectively.The top right panel shows the histogram of the logarithmic flux distribution.
The bottom panel shows the histogram of the logarithmic index distribution. The solid red curve (right-hand top panel and bottom panel) represents the best-fitting function, with dotted green and cyan curves showing the individual components in case of double lognormal fit.}
\label{fig:figa}
\end{figure}

We examined the probability density function (PDF) of the flux and index distribution by constructing a normalized histogram of the logarithm of the flux and spectral index. The histograms were generated with an equal number of points per bin and variable bin widths. The normalized histogram data points are shown in Figure~\ref{fig:figa}, where the corresponding histogram, plotted on a logarithmic scale, is fitted by:
\begin{equation}
D(x) = a \sqrt{\frac{1}{2\pi\sigma_1^2}} \, e^{-\frac{(x-\mu_1)^2}{2\sigma_1^2}} 
+ (1 - a) \sqrt{\frac{1}{2\pi\sigma_2^2}} \, e^{-\frac{(x-\mu_2)^2}{2\sigma_2^2}},
\end{equation}
where $a$ is the normalization fraction, $\mu_1$ and $\mu_2$ are the centroids of the two components, and $\sigma_1$ and $\sigma_2$ are their respective widths of the logarithmic distribution. The fitting resulted in reduced chi-square values of 1.1 and 1.5 for the flux and index distributions, respectively, which are consistent with a double log-normal profile (see Table~\ref{tab:dist}). The double log-normal distribution implies that the source exhibits two distinct flux states. Several studies, including those by \citep{10.1093/mnras/staf620,10.1093/mnras/stz3108, Kushwaha_2016}, have proposed a ``two-flux-state'' hypothesis for blazars, suggesting that these sources can exist in distinct low and high flux states. Our results lend support to this hypothesis, as double log-normal profiles were consistently observed in both the flux and index distributions.

\begin{table*}
    \centering
    \begin{tabular}{ccccccccc}
        \hline
        -- & Distribution      & a    & $\sigma_1$ & $\mu_1$ & $\sigma_2$ & $\mu_2$ & Reduced $\chi^2$ &AD-test statistic\\
        \hline
        Flux  & Double log-normal & $0.56 \pm 0.25$& $0.18 \pm 0.03$    & $-6.75 \pm 0.01 $    & $0.34 \pm 0.05$ & $-6.42 \pm 0.019$  & 1.11&8.6\\
        Index & Double log-normal & $0.18 \pm 0.07$ & $0.02 \pm 0.007$       & $0.36 \pm 0.007  $        & $0.007 \pm 0.003 $     & $0.003 \pm 0.005$ &1.56       & 5.49 \\
        \hline
    \end{tabular}
    \caption{Best-fit parameters obtained from modeling the flux and photon index distributions of PKS\,0402$-$362 with a double log-normal function. 
The parameters $a$, $\sigma_{1,2}$, and $\mu_{1,2}$ represent the relative weight, standard deviations, and mean values of the two log-normal components, respectively. 
The last two columns list the reduced $\chi^{2}$ values and the Anderson–Darling (AD) test statistics.}

    \label{tab:dist}
\end{table*}

\subsection{Search for highest energy photon}

\begin{figure}
\centering
\includegraphics[width=1.0\linewidth]{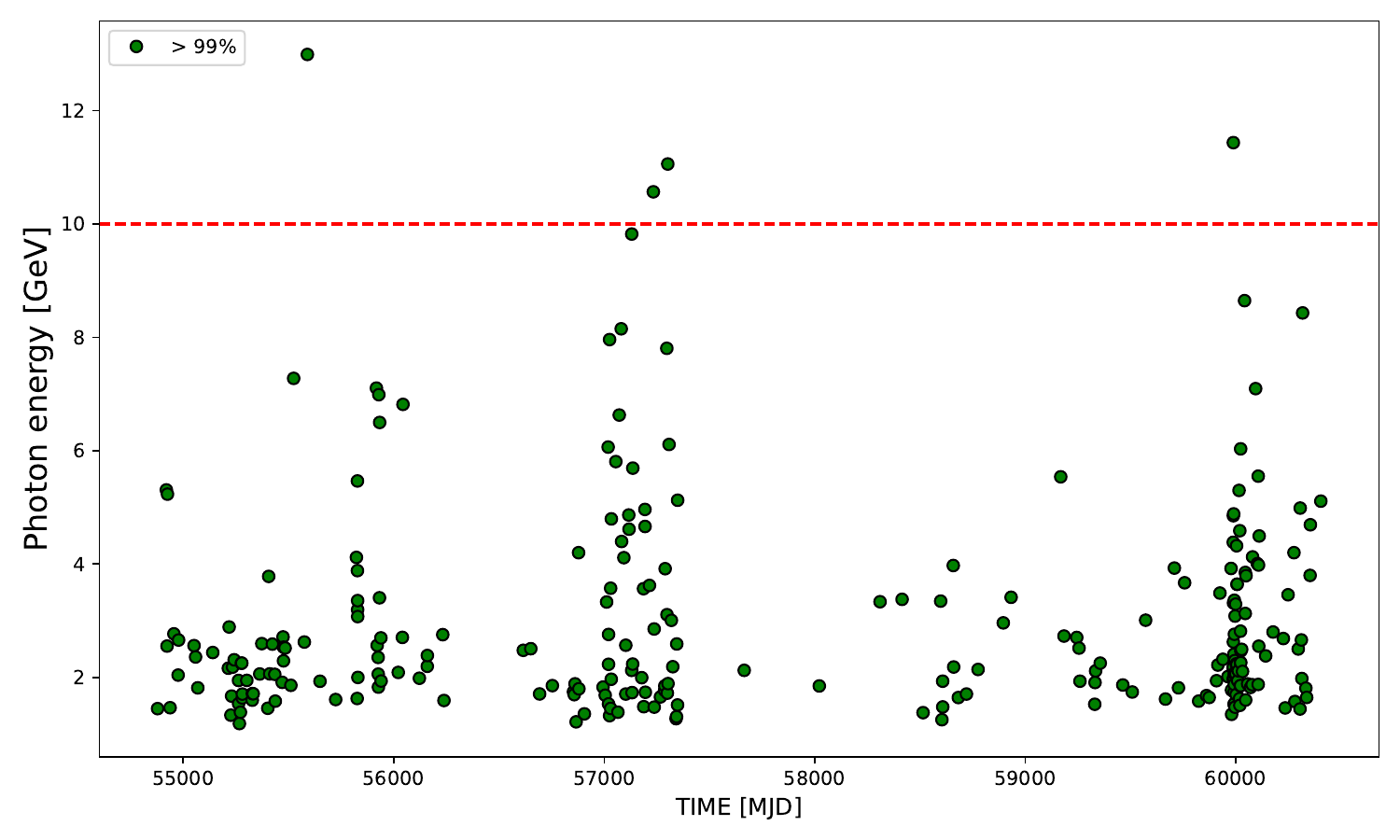}
\caption{Detection of the highest-energy photon from PKS 0402-362}.
\label{fig:high}
\end{figure}

We investigated the emission of high-energy $\gamma$-ray photons with energies exceeding 20~GeV in the rest frame of the source. The detection of such photons is challenging to explain if their origin is assumed to lie within the inner broad-line region (BLR), where photon--photon pair production would likely prevent their escape. To identify high-energy photons potentially associated with PKS~0402$-$362, we employed the \texttt{gtsrcprob} tool from the \textit{Fermi} Science Tools package. This tool computes the probability that detected photons originate from the region of interest (ROI) surrounding the source. We analyzed the light curve of photons with $\geq$99\% probability of association with PKS~0402$-$362 over a time span of approximately 16~years. Our analysis revealed that the highest-energy photon had an energy of 12.9~GeV (see Figure~\ref{fig:high}), which lies below the very-high-energy (VHE) threshold of 20~GeV. Consequently, no VHE photons were identified from PKS~0402$-$362 during this period. This result highlights the difficulty in detecting VHE photons from the source and suggests that future observations with longer integration times or improved sensitivity may be required to probe this energy regime.

%\subsection{Fractional varaibility}

\section{ Search for Quasi-periodic oscillations }\label{sec:qpo}
To investigate the presence of QPOs in the $\gamma$-ray light curve of the blazar PKS~0402$-$362, we employed the Lomb--Scargle Periodogram (LSP) method, along with Monte Carlo simulations to assess the statistical significance of the detected signals. The LSP is a widely used technique for detecting periodic signals in time-series data, particularly when the observations are unevenly sampled \citep{lomb1976least, scargle1982studies}.

In this study, we implemented the LSP using the \textsc{LombScargle}\footnote{\url{https://docs.astropy.org/en/stable/timeseries/lombscargle.html}} function from the \textsc{ASTROPY} library. The measurement uncertainties in the flux values were incorporated to enhance the reliability of the resulting periodogram.
The normalized spectral power $P_{\mathrm{LS}}(f)$ at a frequency $f$ is computed as \citep{vanderplas2018understanding}:
\begin{equation}
\begin{split}
P_{\mathrm{LS}}(f) = \frac{1}{2} \Bigg[ &\frac{\left(\sum_{i=1}^{N} g_i \cos[2\pi f (t_i - \tau)]\right)^2}{\sum_{i=1}^{N} \cos^2[2\pi f (t_i - \tau)]} \\
&+ \frac{\left(\sum_{i=1}^N g_i \sin[2\pi f (t_i - \tau)]\right)^2}{\sum_{i=1}^N \sin^2[2\pi f (t_i - \tau)]} \Bigg],
\end{split}
\end{equation}
where the phase offset $\tau$ is given by
\begin{equation}
\tau = \frac{1}{4\pi f}\tan^{-1}\!\left( \frac{\sum_{i=1}^N \sin(4\pi f t_i)}{\sum_{i=1}^N \cos(4\pi f t_i)} \right).
\end{equation}

\begin{figure*}
\centering
\includegraphics[width=0.9\linewidth]{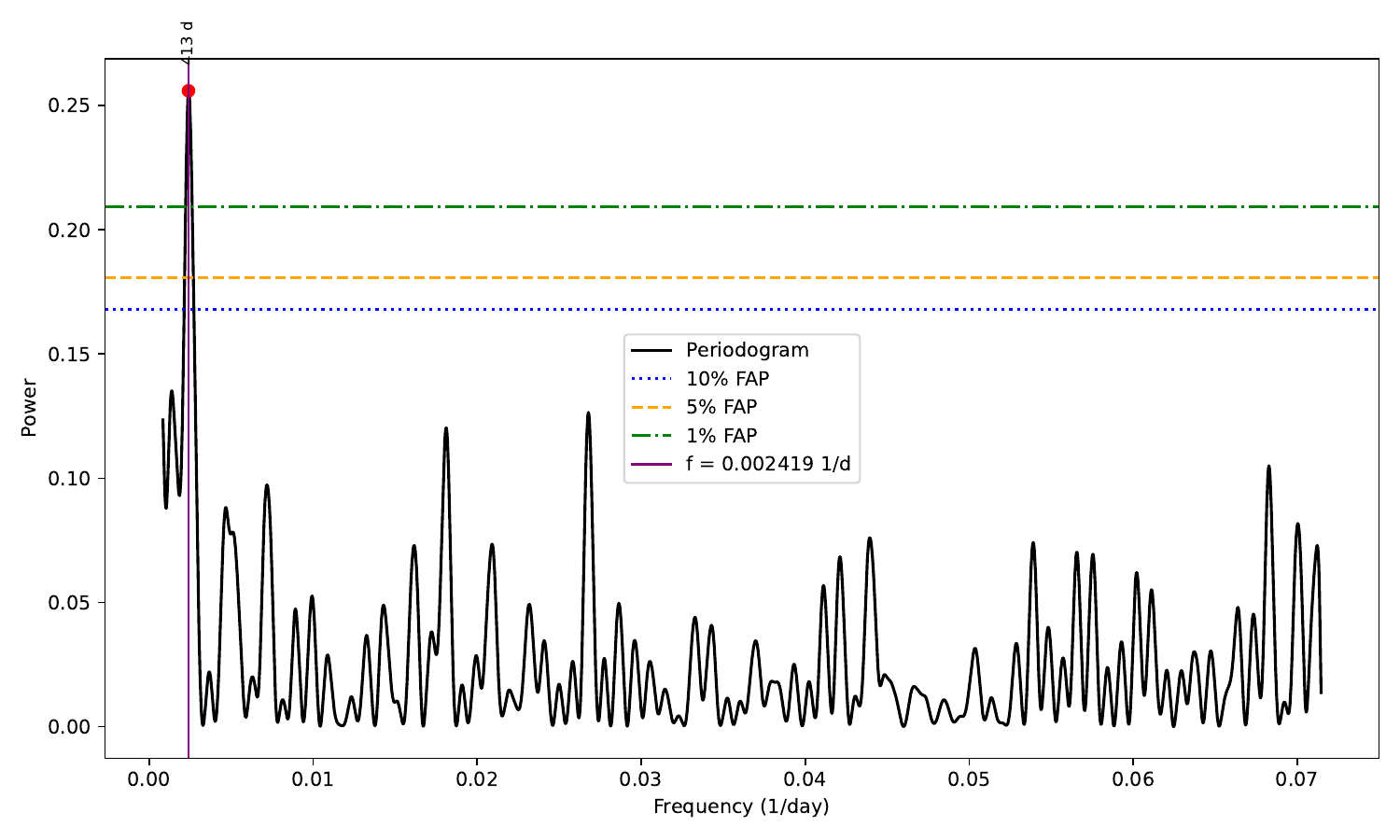}
\caption{Lomb--Scargle periodogram of the $\gamma$-ray light curve of PKS~0402$-$362. The most prominent peak is observed at a frequency of $(24.19 \pm  2.53)\times 10^{-4}$~day$^{-1}$, corresponding to a period of approximately $413.4 \pm 43.2$~days. The horizontal lines represent the false alarm probability (FAP) thresholds at 10\% (blue dotted), 5\% (orange dashed), and 1\% (green dash-dotted) levels, indicating the statistical significance of the detected periodicity.}
\label{fig:lsp}
\end{figure*}

In our analysis, the minimum and maximum search frequencies ($f_{\mathrm{min}}$ and $f_{\mathrm{max}}$) were set to $1/T$ and $1/(2\Delta T)$, respectively, where $T$ is the total observation duration and $\Delta T$ denotes the bin size or mean sampling interval. These limits define the frequency range probed by the LSP. The uncertainty in the detected period was estimated by fitting a Gaussian profile to the main LSP peak and adopting the half-width at half-maximum (HWHM) as the associated error \citep{vanderplas2018understanding, zxgv-fzv5}.
To assess the statistical significance of the peaks detected in the LSP, we performed extensive Monte Carlo simulations. Specifically, we generated $10^5$ synthetic light curves that replicate both the power spectral density (PSD) and the probability density function (PDF) of the observed $\gamma$-ray light curve, following the method proposed by \citet{emmanoulopoulos2013generating}. The local significance of each candidate peak was then evaluated by comparing its power against the distribution of powers at the corresponding frequencies across all simulated datasets.
Additionally, to analytically estimate the false alarm probability (FAP), we used the \texttt{LombScargle.false\_alarm\_probability()} implementation from the \texttt{astropy.timeseries} module with \texttt{method="baluev"}(see Figure \ref{fig:lsp}). This approach is based on the formalism developed by \citet{2008MNRAS.385.1279B}.

\begin{figure*}
\centering
\includegraphics[width=0.9\linewidth]{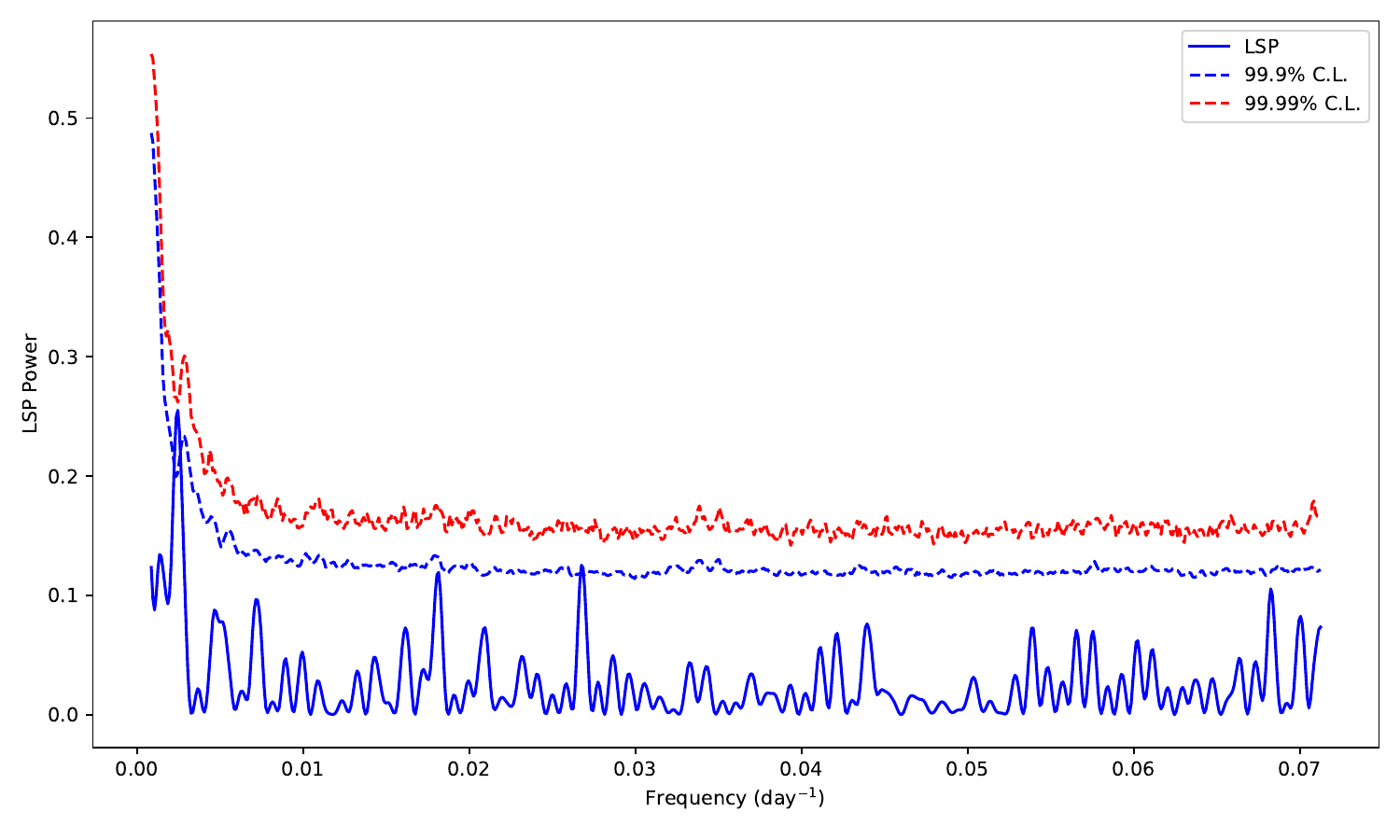}
\caption{Lomb--Scargle periodogram (LSP) of the same light curve, showing a dominant peak at a frequency of 
$0.002419 \pm 0.000253~\mathrm{day^{-1}}$ (corresponding to a period of $\sim413 \pm 43.2~\mathrm{days}$), exceeding the 
$99.9\%$ confidence level derived from $10^{5}$ Monte~Carlo simulations using the method of 
\citet{emmanoulopoulos2013generating}.}
\label{fig:lsp_sim}
\end{figure*}

\begin{figure}
\centering
\includegraphics[width=0.9\linewidth]{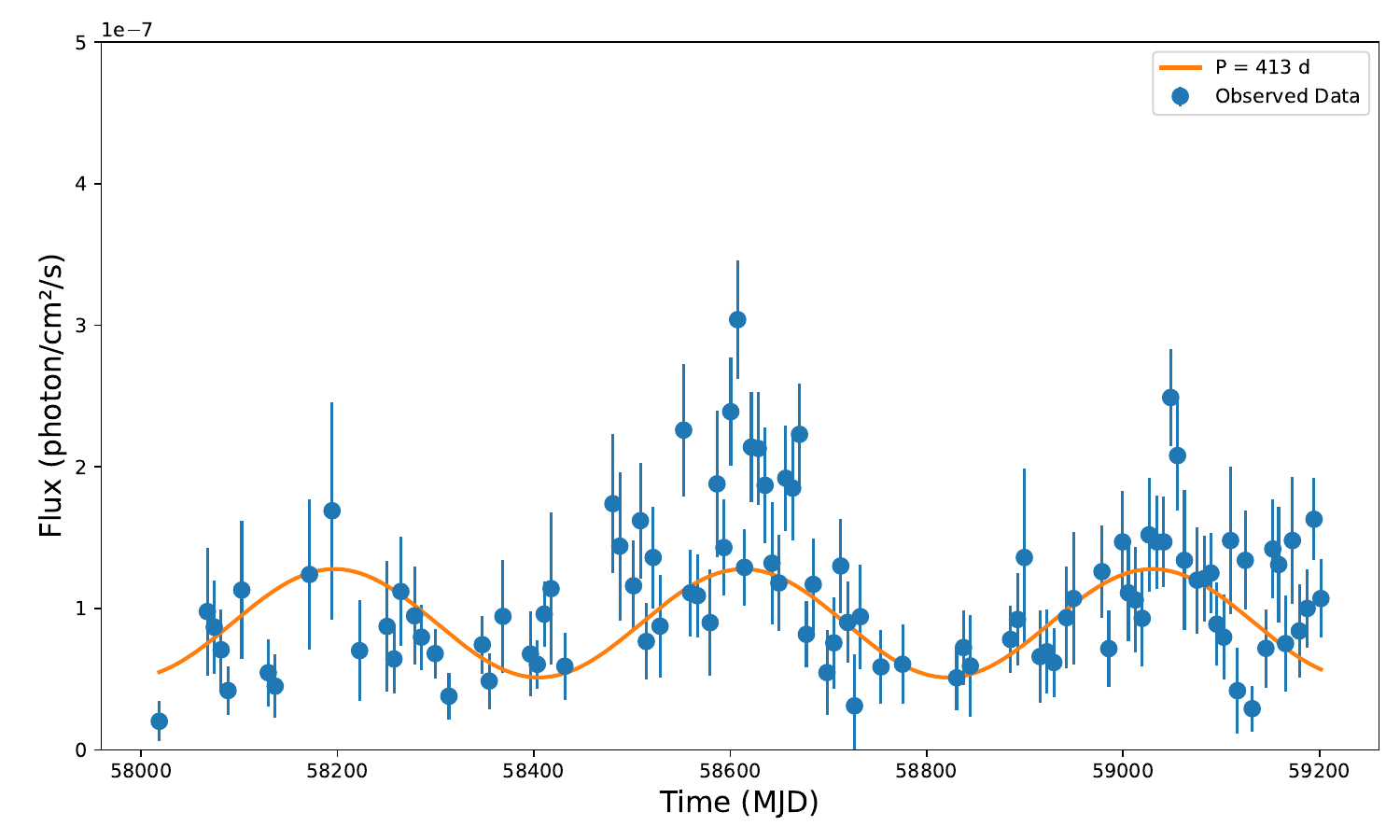}
\caption{Weekly $\gamma$-ray light curve of PKS~0402$-$362 (4FGL~J0403.9$-$3605) spanning MJD~58018--59201, 
showing the best-fit sinusoidal model corresponding to the $\sim$413-day periodicity identified from the Lomb--Scargle periodogram analysis.}
\label{fig:sinefit}
\end{figure}

\begin{figure}
\centering
\includegraphics[width=1.1\linewidth]{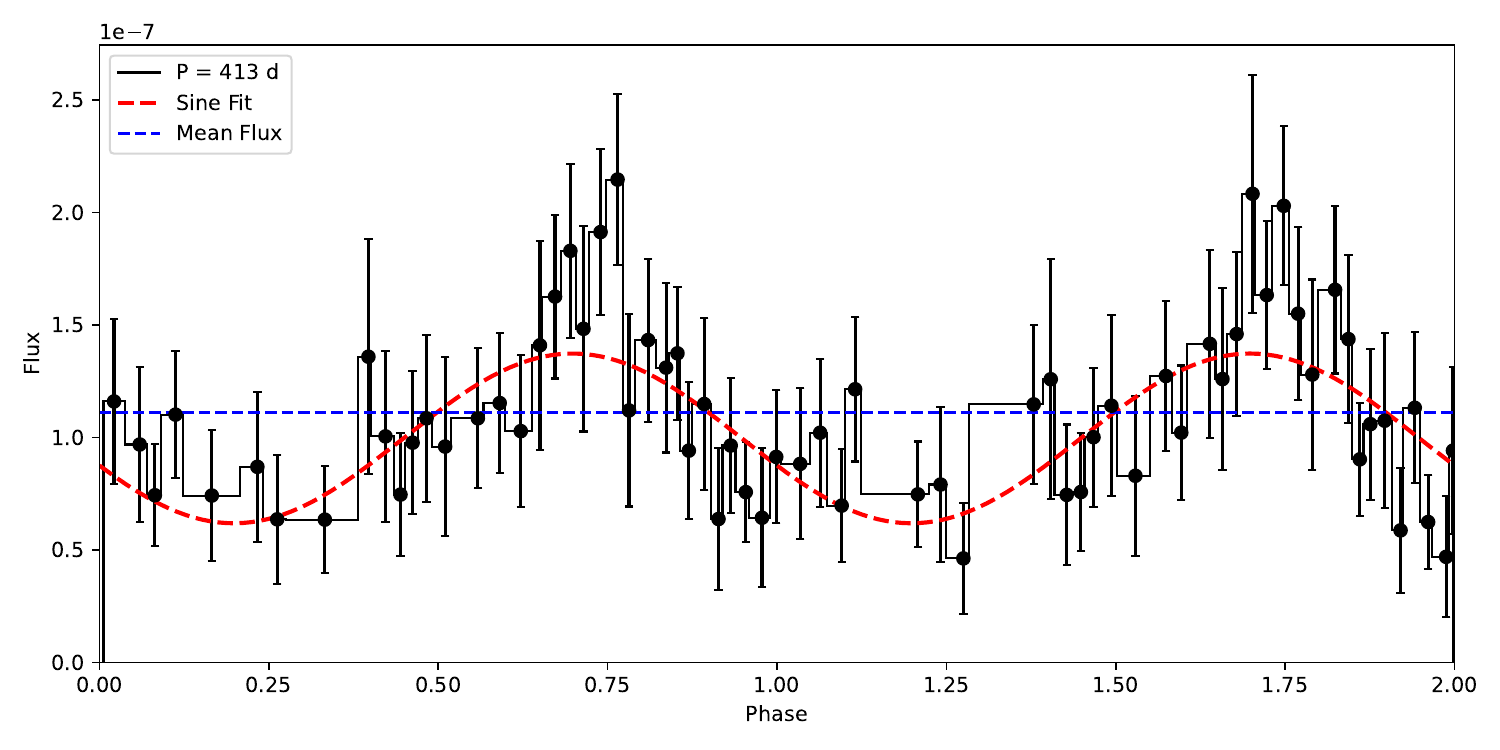}
\caption{. Phase-folded $\gamma$-ray light curves of the blazar PKS 0402-362 for a significant period identified from the Lomb-Scargle periodogram analysis.}.
\label{fig:phase}
\end{figure}

The resulting Lomb--Scargle periodogram is shown in Figure~\ref{fig:lsp_sim}. A prominent peak is detected at a frequency of $(24.19 \pm  2.53)\times 10^{-4}$~day$^{-1}$, corresponding to a period of $413.4 \pm 43.2$~days with a significance level of greater than $3\sigma$. The period corresponds to approximately 3  cycles within the observational window (See  Figure~\ref{fig:sinefit}).  
To further verify the detected periodic behavior, we generated a phase-folded $\gamma$-ray light curve for PKS~0402$-$362 using a period of approximately 413~days and fitted it with a sinusoidal function (see Figure~\ref{fig:phase}). The resulting fit further substantiates the periodic modulation observed in the $\gamma$-ray emission from the source.

However, given that the inferred timescale lies close to one year and spans only $\sim$3 cycles over the available observational window, caution is required in its interpretation. In \textit{Fermi}-LAT observations, annual features can arise from complex and time-dependent variations in spacecraft exposure, solar avoidance geometry, and uneven sampling, rather than from an intrinsic periodic process in the source. We therefore adopt a conservative interpretation and refer to this feature as a \textit{candidate} quasi-periodic modulation.

%---------------- Figure 1: Spectral window ----------------

\section{Broad-band Spectral analysis}\label{sec:sed}
\begin{figure}
\centering
\includegraphics[width=0.7\linewidth,angle=270]{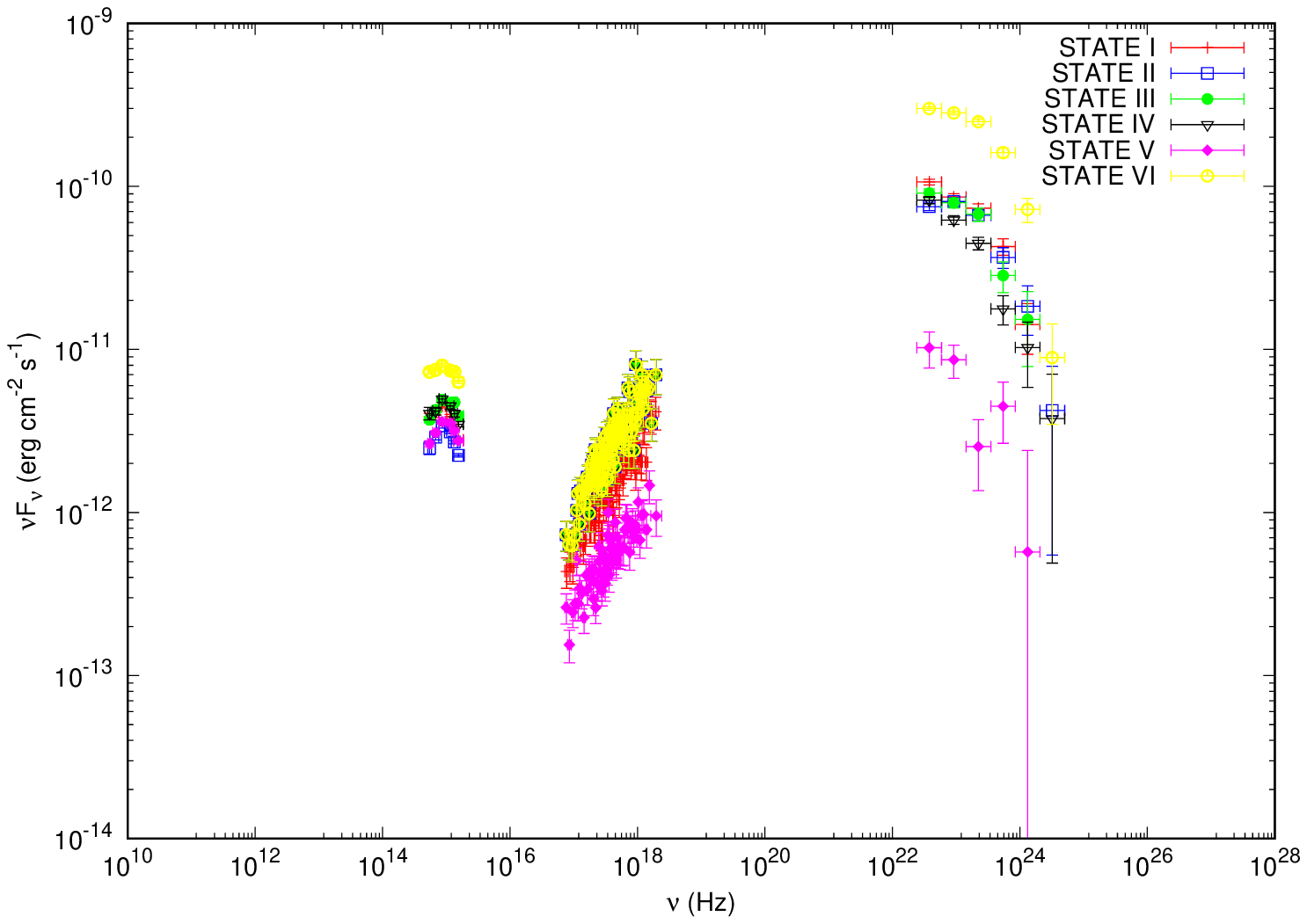}
\vspace{0.6cm}
\caption{The broad-band SED points obtained for S1, S2, S3, S4, S5, and S6 flux states.}
\label{fig:full_sed}
\end{figure}

 To examine the spectral properties and behavior of the source during various flux states, we divided the $\gamma$-ray light curve into segments using BB analysis \citep{Scargle_2013}. We identified six time intervals  with simultaneous $\gamma$-ray, X-ray, and optical/UV observations. These intervals are categorized as the State 1 (S1: MJD 55218-55293), State 2 (S2: MJD 55802-55863), State 3 (S3: MJD 56864-56906), State 4 (S4: MJD 57129-57194), State 5 (S5: MJD 57460-57545) and State 6 (S6: MJD 59974-60025) highlighted by shaded yellow strips in figure \ref{fig:multiplot}. For each flux state under consideration, the $\gamma$-ray SED points are derived by dividing the total energy range of 0.1-300 GeV into eight equally spaced energy bins on a logarithmic scale. During spectral fitting in each energy bin, we assume that the spectral parameters of sources other than PKS 0402-362 within the region of interest (ROI) remain constant with energy, so the parameters for these sources are fixed to their best-fitting values obtained in the 0.1-300 GeV range. The X-ray spectrum for each flux state is generated by providing the specific observation ID to an online automated product generator tool \citep{10.1111/j.1365-2966.2009.14913.x}. The source spectra are then binned using the GRPPHA task to ensure 20 counts per bin, and the resulting spectra are fitted with an absorbed power-law model. For Swift-UVOT, the images corresponding to the observation IDs within each flux state are combined using the UVOTIMSUM task, and the optical/UV flux values are extracted from the combined image. The broad-band SED points obtained for the S1, S2, S3, S4, S5, and S6 flux states are presented in Figure \ref{fig:full_sed}.

The broad-band SEDs for different flux states were modeled using a one-zone leptonic emission framework \citep{10.1093/mnras/stx1194, 2018RAA....18...35S}. The radiation is assumed to originate from a homogeneous spherical region of radius $R$ embedded within a relativistic jet that moves with a bulk Lorentz factor $\Gamma$. The jet is inclined at a small angle $\theta$ to the observer’s line of sight, resulting in Doppler boosting of the emitted flux. The Doppler factor is defined as $\delta = [\Gamma (1 - \beta \cos \theta)]^{-1}$.
We express the electron Lorentz factor $\gamma$ in terms of a dimensionless variable $\xi$, defined as $\xi = \gamma \sqrt{\mathbb{C}}$, where $\mathbb{C} = 1.36 \times 10^{-11}\, \delta B / (1 + z)$ \citep{AKBAR2025438,10.1093/mnras/stab1244}. The emitting region is populated with relativistic electrons following a broken power-law (BPL) energy distribution, given by

\begin{eqnarray}
n(\xi)\, d\xi =
\begin{cases}
K\, \xi^{-p}\, d\xi, & \text{for } \xi_{\mathrm{min}} < \xi < \xi_{\mathrm{b}}, \\
K\, \xi_{\mathrm{b}}^{q-p}\, \xi^{-q}\, d\xi, & \text{for } \xi_{\mathrm{b}} < \xi < \xi_{\mathrm{max}},
\end{cases}
\end{eqnarray}
where $K$ is the normalization constant, $\xi_{\mathrm{b}}$ is the break energy, and $p$ and $q$ are the spectral indices below and above the break, respectively. The parameters $\xi_{\mathrm{min}}$ and $\xi_{\mathrm{max}}$ denote the minimum and maximum energies of the electrons.
Relativistic electrons interacting with the magnetic field $B$ and the surrounding photon field radiate through synchrotron and IC processes. In this analysis, the seed photons for the IC scattering are assumed to be the synchrotron photons produced within the jet itself, giving rise to SSC emission.
 Following \citet{1984RvMP...56..255B, 1995ApJ...446L..63D, 2008ApJ...686..181F, 2018RAA....18...35S}, the synchrotron flux at photon energy $\epsilon$ can be written as:
\begin{equation}\label{eq:syn_flux}
F_{\mathrm{syn}}(\epsilon)=\frac{\delta^3(1+z)}{d_L^2}\,V\,\mathbb{A}\int_{\xi_{\mathrm{min}}}^{\xi_{\mathrm{max}}} f(\epsilon/\xi^2)\,n(\xi)\,d\xi,
\end{equation}

where $d_L$ is the luminosity distance, $V$ is the emitting volume, and $\mathbb{A} = \frac{\sqrt{3}\pi e^3 B}{16m_e c^2 \sqrt{\mathbb{C}}}$. The quantities $\xi_{\mathrm{min}}$ and $\xi_{\mathrm{max}}$ correspond to the minimum and maximum electron energies, respectively, and $f(x)$ denotes the synchrotron emissivity function as defined in \citet{1986rpa..book.....R}.

The observed SSC flux at energy $\epsilon$ is expressed as \citep{2008ApJ...686..181F, 2018RAA....18...35S}:

\begin{equation}\label{eq:ssc_flux}
\begin{split}
F_{\mathrm{ssc}}(\epsilon) = \frac{\delta^3(1+z)}{d_L^2}\,V\,\mathbb{B}\,\epsilon &\int_{\xi_{\mathrm{min}}}^{\xi_{\mathrm{max}}} \frac{1}{\xi^2} 
\int_{x_1}^{x_2} \frac{I_{\mathrm{syn}}(\epsilon_i)}{\epsilon_i^2} \\
&\times f(\epsilon_i, \epsilon, \xi/\sqrt{\mathbb{C}})\,d\epsilon_i\,n(\xi)\,d\xi,
\end{split}
\end{equation}

where $\epsilon_i$ is the energy of the seed photon, $\mathbb{B} = \frac{3}{4}\sigma_T\sqrt{\mathbb{C}}$, and $I_{\mathrm{syn}}(\epsilon_i)$ is the synchrotron intensity. The integration limits are defined as 
$x_1 = \frac{\mathbb{C}\,\epsilon}{4\xi^2(1-\sqrt{\mathbb{C}}\,\epsilon/\xi m_e c^2)}$ and 
$x_2 = \frac{\epsilon}{(1-\sqrt{\mathbb{C}}\,\epsilon/\xi m_e c^2)}$. 
The scattering kernel is given by

\begin{equation}
f(\epsilon_i, \epsilon, \xi) = 2q\ln q + (1+2q)(1-q) + \frac{\kappa^2 q^2 (1-q)}{2(1+\kappa q)}, 
\end{equation}

where $q = \frac{\mathbb{C}\epsilon}{4\xi^2\epsilon_i(1-\sqrt{\mathbb{C}}\epsilon/\xi m_e c^2)}$ and $\kappa = \frac{4\xi\epsilon_i}{\sqrt{\mathbb{C}}m_e c^2}$.

Similarly, the EC flux at energy $\epsilon$ can be obtained as \citep{2018RAA....18...35S}:

\begin{equation}\label{eq:ec_flux}
\begin{split}
F_{\mathrm{ec}}(\epsilon) = \frac{\delta^3(1+z)}{d_L^2}\,V\,\mathbb{D}\,\epsilon
\int_{0}^{\infty} d\epsilon_i^* \int_{\xi_{\mathrm{min}}}^{\xi_{\mathrm{max}}} \frac{N(\xi)}{\xi^2}
\frac{U_{\mathrm{ph}}^*}{\epsilon_i^*} \eta(\xi,\epsilon_s,\epsilon_i')\,d\xi,
\end{split}
\end{equation}

where $\mathbb{D} = \frac{3}{32\pi} c \beta \sigma_T \sqrt{\mathbb{C}}$, $\epsilon_i^*$ represents the energy of target photons in the AGN frame, and $U_{\mathrm{ph}}^*$ is their energy density. The function $\eta(\xi,\epsilon_s,\epsilon_i')$ is defined as

\begin{equation}
\eta(\xi,\epsilon_s,\epsilon_i') = y + \frac{1}{y} + \frac{\mathbb{C}\epsilon_s^2}{\xi^2\epsilon_i'^2y^2} - \frac{2\nu_s\sqrt{\mathbb{C}}}{\xi\epsilon_i'y},
\end{equation}
with $y = 1 - \frac{\sqrt{\mathbb{C}}\epsilon_s}{\xi m_e c^2}$.

Equations~\ref{eq:syn_flux}, \ref{eq:ssc_flux}, and \ref{eq:ec_flux} were solved numerically for a broken power-law electron energy distribution. The resulting numerical code was implemented as a local convolution model in \textsc{XSPEC} for broadband SED fitting.  The observed broadband spectrum is governed by ten key parameters: $\xi_{\mathrm{b}}$, $\xi_{\mathrm{min}}$, $\xi_{\mathrm{max}}$, $p$, $q$, $\Gamma$, $B$, $R$, $\theta$, and the normalization constant $N$. The model framework also permits fitting with the jet power ($P_{\mathrm{jet}}$) as a free parameter while keeping $N$ fixed. The initial parameter values were estimated from the approximate shape and flux levels of the synchrotron, SSC, and EC components during the flaring state.

To account for possible model deviations and residual thermal contamination, a systematic uncertainty of 13\% was added to the UVOT flux points. This systematic ensures that the optical/UV spectrum can be well described by a single power-law component, consistent with the assumption that the emission in this band is predominantly non-thermal (synchrotron) in origin. A similar approach has been adopted in earlier blazar SED modeling studies \citep[e.g.][]{10.1093/mnras/stae588,10.1093/mnras/stad3818}. The inclusion of this systematic term effectively compensates for any additional contribution from external thermal components such as the accretion disc, which may otherwise cause mild deviations from the pure power-law behavior in the optical/UV regime. Such a thermal excess from the accretion disc in PKS\,0402$-$362 has also been reported by \citet{2023MNRAS.521.3451D}, supporting our analysis procedure. The introduction of this systematic uncertainty was further necessary to achieve acceptable fits, yielding reduced $\chi^2$ values typically below 2 for most of the SED states.

\begin{figure*}
    \begin{center}
        % First row
        \includegraphics[angle=270,width=.47\textwidth]{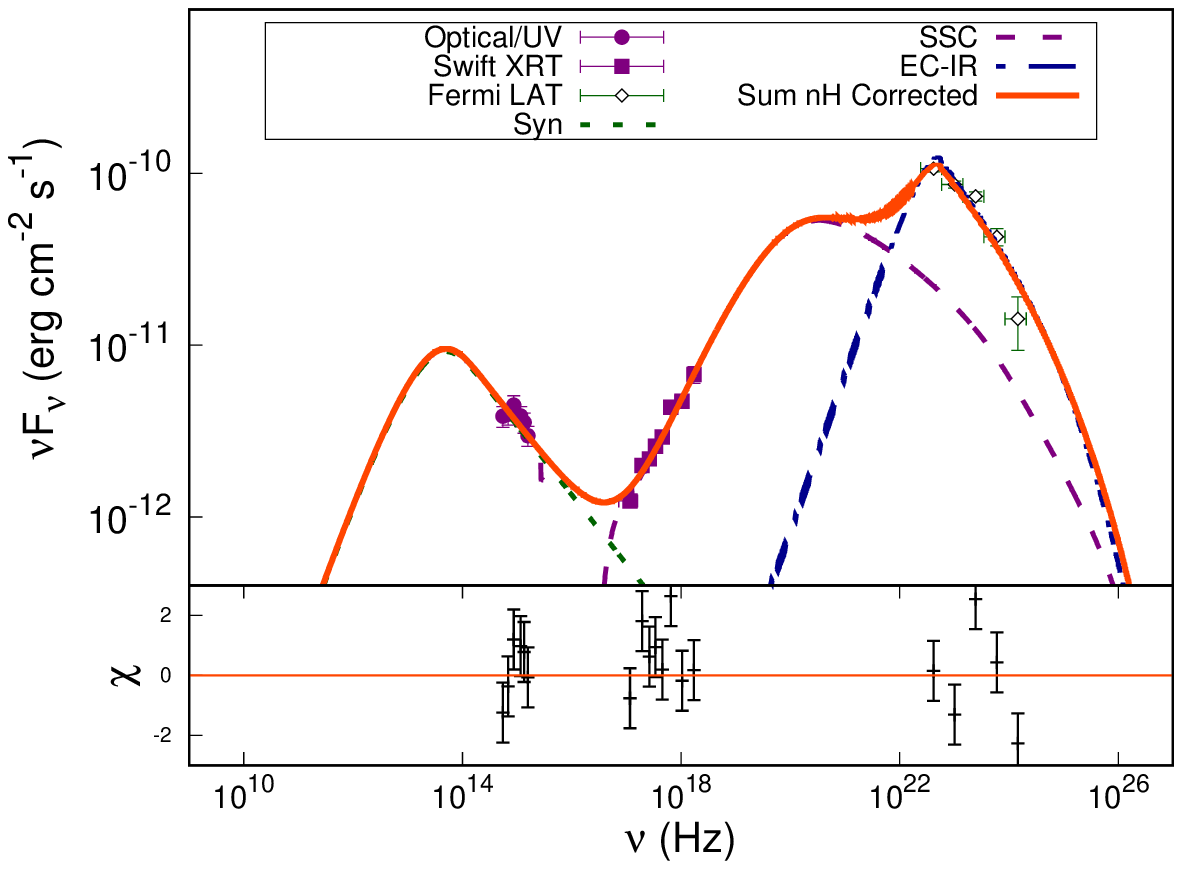}
        \includegraphics[angle=270,width=.47\textwidth]{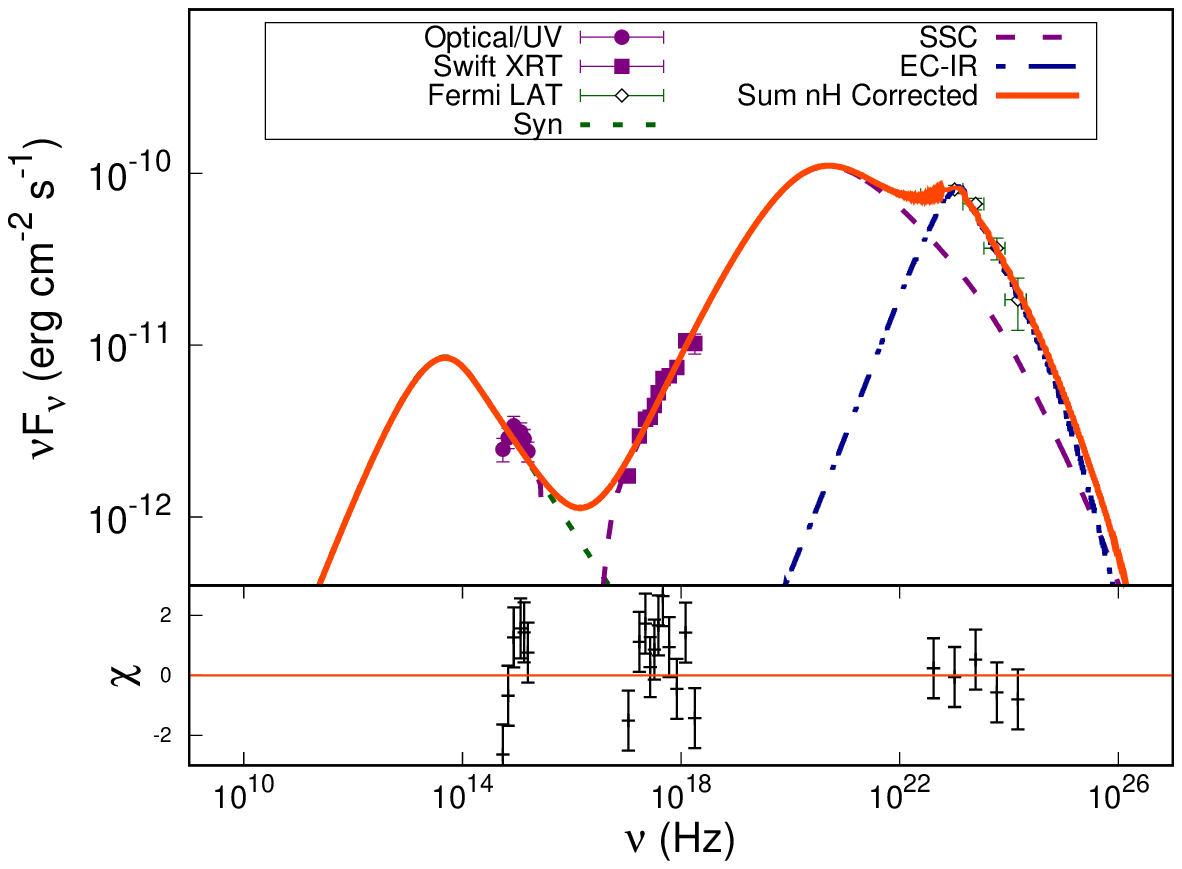}\\[6pt]
        
        % Second row
        \includegraphics[angle=270,width=.47\textwidth]{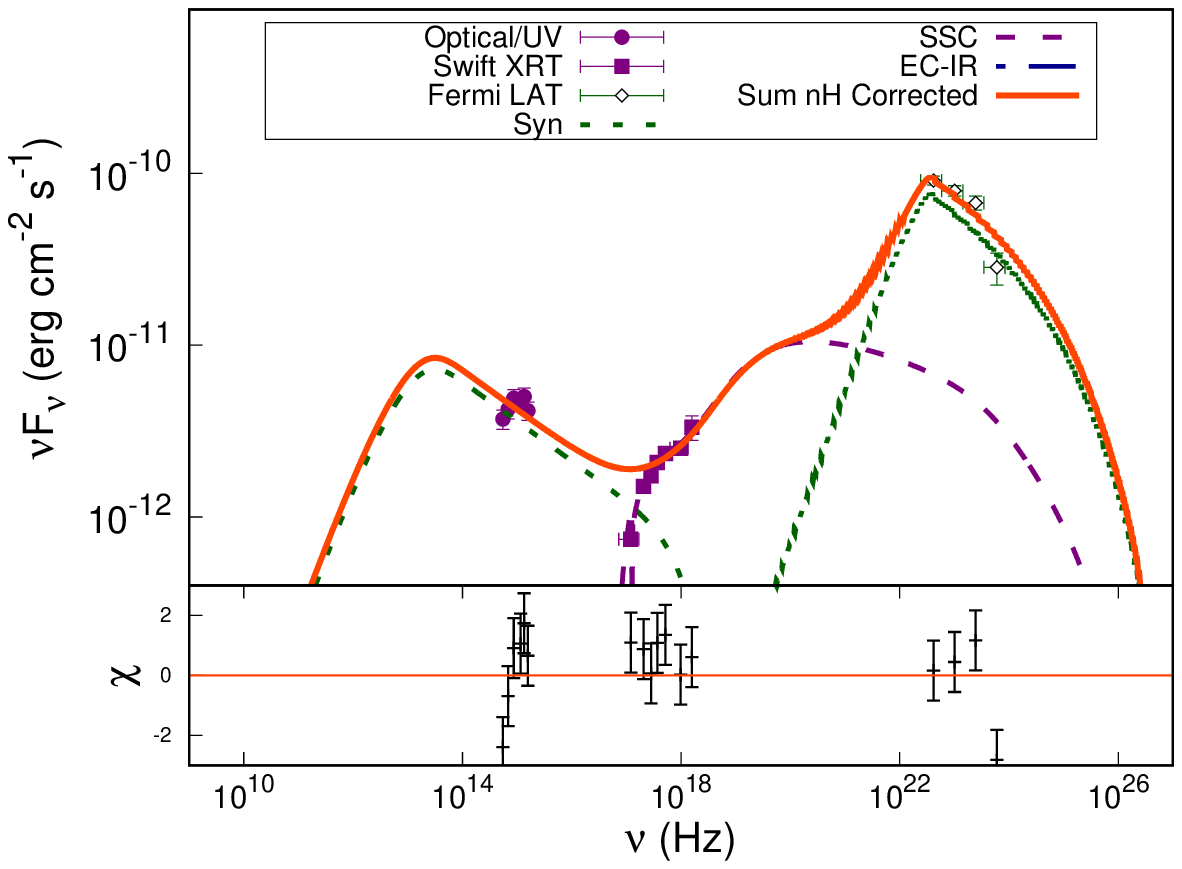}
        \includegraphics[angle=270,width=.47\textwidth]{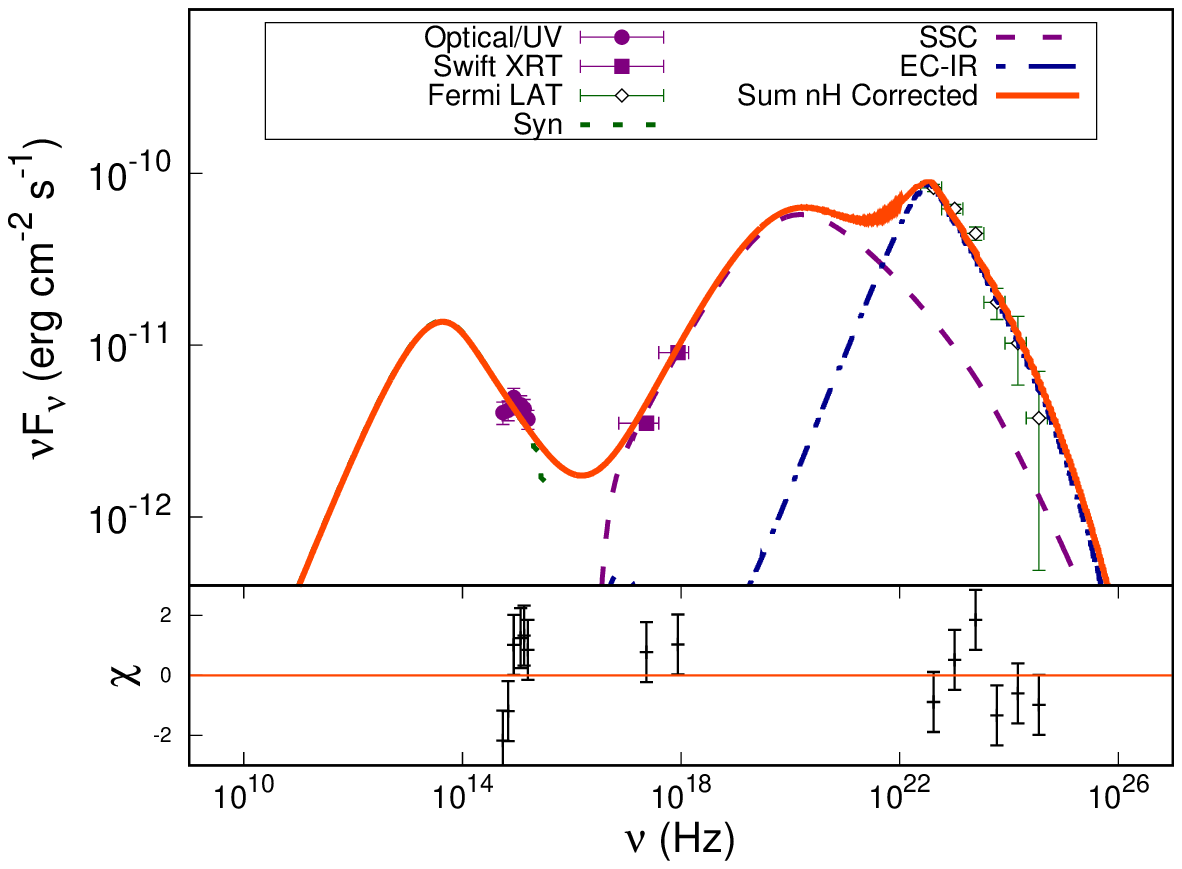}\\[6pt]
        
        % Third row
        \includegraphics[angle=270,width=.45\textwidth]{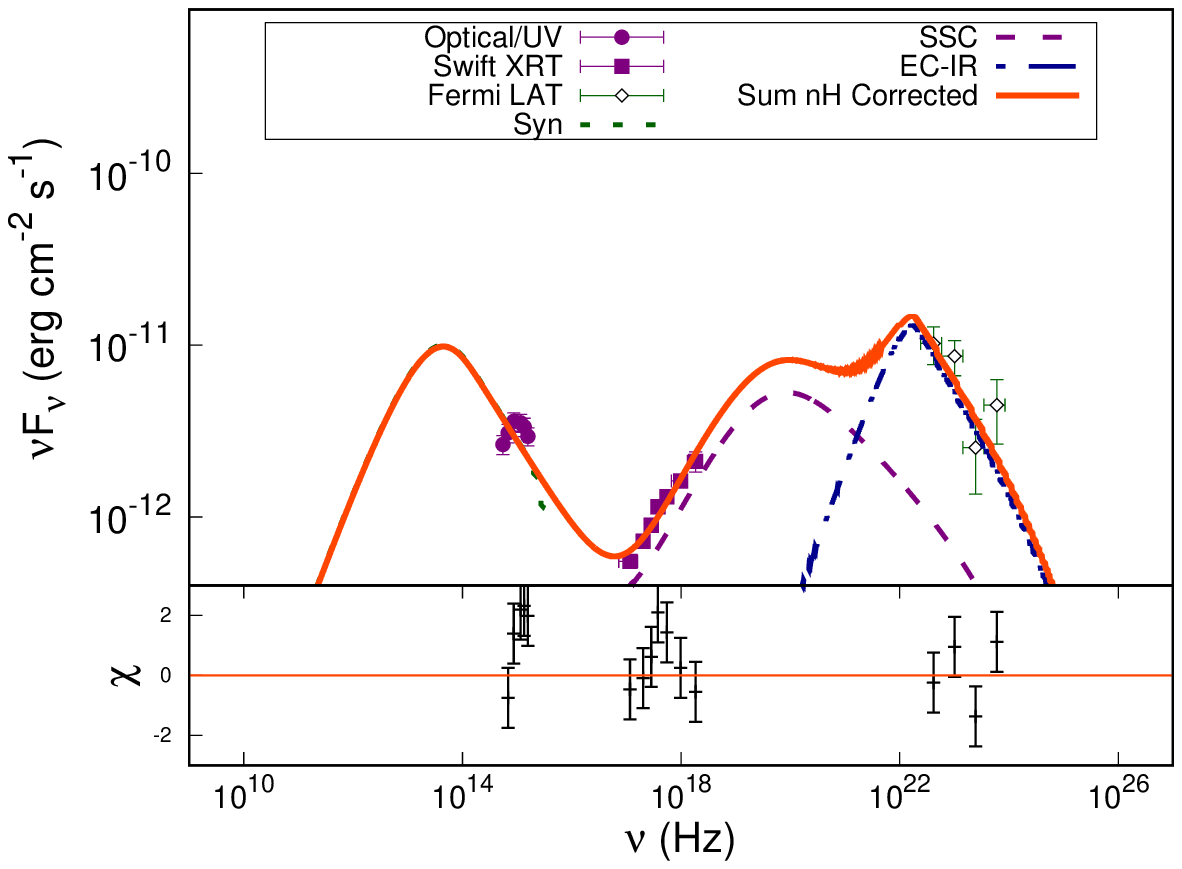}
        \includegraphics[angle=270,width=.47\textwidth]{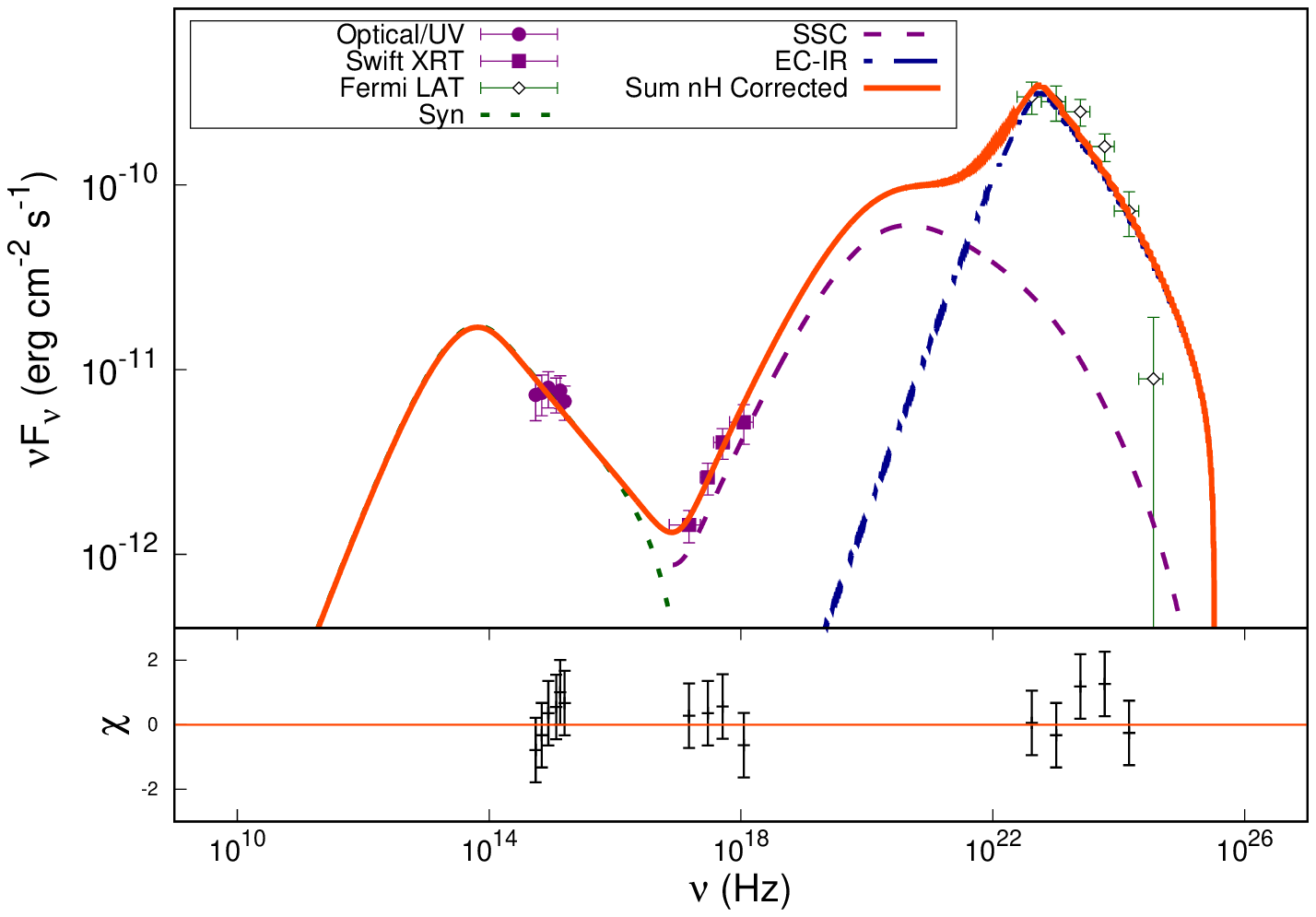}
        \vspace{0.9cm}
        
        \caption{Broadband SEDs of PKS\,0402$-$362 during flux states S1--S6. 
        The flux points denoted by filled diamonds correspond to \textit{Fermi}-LAT, 
        squares correspond to \textit{Swift}-XRT, and filled circles correspond to \textit{Swift}-UVOT. 
        The solid red curve illustrates the combined best-fit synchrotron, SSC, and EC spectra.}
        \label{fig:sed_all}
    \end{center}
\end{figure*}

 Owing to limited observational coverage in the $\gamma$-ray, X-ray, and optical/UV bands, only four parameters — $p$, $q$, $B$, and $\Gamma$ — were allowed to vary freely during the fitting process, while the remaining parameters were fixed at their representative best-fit values. 
The observed spectral data points, along with the modeled broadband SED and individual contributions from the synchrotron, SSC, and EC components, are presented in Figure~\ref{fig:sed_all}. The corresponding best-fitting parameters are listed in Table~\ref{tab:sed_res}. We found that the synchrotron, SSC, and EC components together provided satisfactory fits to the observed spectra across all flux states.

\begin{table*}
    \centering
    \caption{\label{tab:sed_res}Best-fit model parameters for the broadband SEDs of PKS\,0402-362 in the quiescent and flaring states. 
The table is organized into two sections: free parameters and fixed parameters.
The top section lists the free parameters varied during the fit namely: broken power law indices of the electron distribution before and after the break energy, $p$ and $q$, Bulk Lorentz factor of the electron energy distribution ($\Gamma_{b}$) and the magnetic field strength ($B$), expressed in units of Gauss. 
The middle section includes fixed parameters used in the modeling: $\xi_{\rm b}$ and $\xi_{\max}$ represent maximum and break electron energies, respectively, with all energies expressed in units of $\sqrt{\mathrm{keV}}$. In all the flux states, we fixed the  minimum electron energy ($\xi_{min}$) at $10^{-6}$ $\sqrt{\mathrm{keV}}$, size of emission region at $1.32 \times 10^{16}$ cm and  jet inclination angle at ($\theta = 2$ degree).
The bottom section presents the logarithm of the jet power (erg\,s$^{-1}$)) and the reduced $\chi$-square value ($\chi^{2}/\mathrm{dof}$) from the spectral fit   .    
    }
    \begin{tabular}{lccccccc} \hline
Free Parameter   & Symbol &  State I  & State II  & State III & State IV & State V & State VI \\ \hline

Low energy spectral index     & p  &  $1.156^{1.161}_{1.147}$ &     $1.33^{1.35}_{1.34}$     & $1.2^{1.22}_{1.20}$     & $1.40^{ 1.41}_{1.39}$ & $1.22^{1.23}_{1.21}$ & $1.138^{1.16}_{1.13}$ \\
High energy spectral index      & q   &    $3.8^{3.89}_{3.74}$ &   $3.9^{4.03}_{3.86}$  & $3.45^{3.49}_{3.41}$     &   $3.96^{4.04}_{3.9}$     &     $4.01^{4.14}_{3.9}$     &$3.82^{3.9}_{3.6}$   \\
Bulk Lorentz factor                     & $\Gamma$    & $48.3^{--}_{42.1}$
               &    $31.5^{45.1}_{18.1}$ & $41.03^{--}_{28.3}$ & $50^{--}_{44.2}$ & $49.85^{--}_{42.6}$ &  $38.4^{--}_{15.8}$ \\
Magnetic field                  & B    &  $0.46 ^{0.5}_{0.42}$  &  $0.22^{0.23}_{0.20}$   &  $0.47^{0.49}_{0.43}$ &    $0.6 ^{0.65}_{0.58}$  &  $1.26 ^{1.42}_{1.13}$ &  $0.52 ^{0.6}_{0.5}$  \\
\hline
Fixed Parameters\\
\hline
High energy Lorentz factor $\sqrt{\mathrm{keV}}$ & $\xi_{\rm max}$ &  10& 10& 2& 10& 10&  0.41\\
Break Lorentz factor$ \sqrt{\mathrm{keV}}$ & $\xi_{\rm b}$  &    $1.14\times 10^{-2}$ &  $1.19\times 10^{-2}$  & $8.11\times 10^{-3}$   & $1.14\times 10^{-2}$ &  $1.14\times 10^{-2}$  &  $1.31\times 10^{-2}$    \\
\hline 
Jet powet& $\rm \log_{10}{P_{jet}}$  & 46.41 & 47.73 & 46.96 & 47.82 & 45.49 & 47 \\
Reduced chi-squared & $\chi^2/d.o.f$  &105.64/85 &135.77/109 &88.13/65&38.5/27&101.86/71 & 25.91/23\\

\hline
\end{tabular}
\end{table*}

\section{SUMMARY AND DISCUSSION}

We conducted a detailed multiwavelength investigation of FSRQ PKS~0402$-$362 using data from Fermi-LAT and Swift-XRT/UVOT, spanning the period from MJD 54686-60321. This comprehensive analysis combined temporal, statistical, and spectral methods to probe the variability behaviour and underlying emission mechanisms of the source.
The long-term $\gamma$-ray light curve revealed several pronounced flaring episodes, including a major outburst on 2023 February 13, with a daily averaged $\gamma$-ray flux (E>100 MeV) of $(3.5 \pm 0.2) \times 10^{-6} \, \mathrm{ph \, cm^{-2} \, s^{-1}}$. This corresponds to a flux increase of factor 20 relative to the average flux reported in the fourth Fermi-LAT catalog (4GFL-DR3) \citep{Abdollahi_2022}. A moderate ``harder-when-brighter'' trend between the photon index and $\gamma$-ray flux was observed, indicating that the IC peak shifts toward higher energies during high states \citep{10.1093/mnras/stz151} .
The fractional variability ($F_{\mathrm{var}}$) analysis across multiple energy bands revealed that the variability amplitude is highest in $\gamma$-rays and lowest in X-rays, with intermediate values in the optical/UV bands. This pattern suggests that the most energetic electrons, responsible for $\gamma$-ray emission, are more sensitive to radiative losses and acceleration variations, leading to stronger flux fluctuations. The X-ray emission likely represents the low-energy tail of the Compton component, while the optical/UV and $\gamma$-ray emissions trace the high-energy ends of the synchrotron and Compton components, respectively.
The variability dependence on time binning showed that when all data points were included, $F_{\mathrm{var}}$ increased from the 3-day to the 7-day bin and slightly decreased at 30 days. However, after applying a flux significance cut ($\mathrm{flux}/\mathrm{flux_{error}} > 3$), $F_{\mathrm{var}}$ exhibited a steady rise with increasing bin size, implying that the intrinsic variability amplitude is more pronounced on longer timescales. This behavior is consistent with previous studies of FSRQs \citep{2025PhRvD.111l3052S, Akbar_2024}, which reported a gradual increase in $F_{\mathrm{var}}$ with bin size, whereas BL~Lac objects generally show a decrease \citep{2019Galax...7...62S}.

The statistical analysis of flux and spectral index distributions provides insight into the underlying variability mechanisms of PKS~0402$-$362. Using the AD test and histogram fitting, we obtained test statistic values of 8.6 and 4.9 for the flux and index distributions, respectively, both exceeding the critical value at 5\% significance threshold. This indicates that neither distribution follows a single normal or log-normal profile. Instead the histogram fitting suggest that data are better represented by a double log-normal function, consistent with a bimodal or “two-state” behaviour. Such behaviour suggests that the source alternates between distinct low and high flux states, a pattern also reported in earlier blazar studies \citep{10.1093/mnras/staf620,10.1093/mnras/stz3108, 2018RAA....18..141S, Kushwaha_2016}. The presence of two flux states could reflect transitions between quiescent and active emission phases, possibly driven by changes in particle acceleration efficiency or variations in the jet environment.

We also investigated the presence of VHE $\gamma$-ray photons from PKS~0402$-$362 to examine possible emission beyond the conventional Fermi-LAT energy range. Using the \texttt{gtsrcprob} tool within the \textit{Fermi} Science Tools, we analyzed photons with $\geq$99\% probability of association with the source over a 16-year period. The highest-energy photon detected had an energy of 12.9~GeV (see Figure~\ref{fig:high}), which falls below the VHE threshold of 20\,GeV. 
%The absence of photons above this limit implies that $\gamma$-rays likely originate from regions within or near the BLR, where pair production processes can attenuate photons above tens of GeV. 
This finding is consistent with previous observations of FSRQs, where internal absorption and external Compton dominance suppress VHE emission. Future observations with improved sensitivity or longer integration times, particularly from facilities such as the Cherenkov Telescope Array (CTA), will be essential to probe potential VHE components and constrain the location of the $\gamma$-ray emission zone.

A modulation on a timescale of $413 \pm 43$~days, exceeding the 99.9\% confidence level based on $10^5$ Monte Carlo simulations, is indicated in the $\gamma$-ray light curve of PKS~0402$-$362 through Lomb--Scargle analysis. However, as discussed in Section~\ref{sec:qpo}, the $\sim$413-day feature lies close to the annual timescale and is supported by only a limited number of cycles, warranting a cautious interpretation. Accordingly, we treat this signal conservatively as a \textit{candidate} quasi-periodic modulation rather than evidence for a firmly established intrinsic periodicity.

The physical origin of quasi-periodic variability in AGN remains an open problem, and a variety of mechanisms have been proposed in the literature to explain long-term modulations when such signals are established as intrinsic. Supermassive binary black hole (SMBBH) systems have been invoked to explain multi-year periodicities, most notably in OJ~287 \citep{1996ApJ...460..207L,1997ApJ...484..180S,2006ApJ...643L...9V}. However, the orbital timescales expected for genuine binary SMBH systems typically span several years to decades \citep{2007Ap&SS.309..271R}, making a $\sim$400-day period difficult to reconcile with simple binary orbital motion.
Alternative scenarios that can produce shorter characteristic timescales include geometric effects such as continuous jet precession or Lense–Thirring precession of a misaligned accretion disk \citep{romero2000beaming, rieger2004geometrical, stella1997lense, liska2018formation}. In such models, modest changes in jet orientation can lead to quasi-periodic Doppler boosting of the observed emission without requiring strictly periodic intrinsic power variations. Other proposed mechanisms include orbiting inhomogeneities or hotspots in the inner accretion flow that modulate the seed photon field for external Compton scattering \citep{zhang1990rotation, gupta2008periodic}, as well as episodic magnetic reconnection events within the jet that may generate recurrent flaring activity \citep{huang2013magnetic}.

 Broadband spectral energy distributions for six flux states of PKS~0402$-$362 were successfully modeled within a one-zone leptonic framework incorporating synchrotron, SSC, and EC components. The low-energy hump is well reproduced by synchrotron emission, while the high-energy component requires both SSC and EC contributions. For the EC process, infrared photons from a dusty torus (approximated as a blackbody with $T \sim 1000$\,K) provide satisfactory fits to the \textit{Fermi}-LAT data in the GeV band. These SED modeling results are consistent with  typically  powerful FSRQs \citep{2021MNRAS.504..416S, 2022MNRAS.514.4259M}. However, in some FSRQs the $\gamma$-ray emission also requires seed photons from the BLR for the IC process. For example, \citet{2022MNRAS.513..611K} showed that while the quiescent state can be adequately reproduced using only EC/IR emission, flaring states require a combination of EC/IR and EC/BLR components. Such results imply that, in different sources, the emission region may lie either inside or outside the BLR. In contrast, for PKS~0402$-$362 our modeling consistently places the emission region outside the BLR in both low and high flux states. The location of the emission region beyond the BLR, together with the use of IR seed photons, is supported by the detection of several $\geq 10$\,GeV photons with association probability $>99.99\%$.  Otherwise  $\gamma\gamma$ absorption inside the  BLR would suppress high energy $\gamma$-ray photons,  if the emission originated within the BLR \citep[e.g.,][]{2011A&A...534A..86T}.
 
The broadband SED identifies S6 as the highest-flux state and S5 as the lowest, with the remaining states lying in between (see Figure \ref{fig:sed_all}). The best-fit SED parameters show systematic trends between low- and high-flux states: lower-flux epochs are characterized by larger magnetic fields ($B \sim 1.26$ G) and steeper injected electron spectra (higher values of $p$ and $q$), whereas brighter states exhibit spectral hardening and a decrease in $B$. Overall, the magnetic field strength ranges from $0.22$ to $1.26$ G. A similar trend of higher $B$ during low flux states has been reported previously for this source \citep{2023MNRAS.521.3451D}.
The reduction in magnetic energy density during high-flux epochs implies that the jet becomes more particle- or kinetic-energy–dominated in its active phases, leading to an increase in Compton dominance. Such behavior has been commonly observed in powerful FSRQs and can be interpreted as resulting from enhanced particle acceleration and/or decreased magnetization during flaring episodes \citep[e.g.,][]{2009MNRAS.397..985G}. Our fitted electron distributions are described by a broken power law; however, the observed difference between the low- and high-energy indices ($\Delta p = p_2 - p_1$) exceeds the canonical value of unity expected for a simple radiative cooling break. This suggests that processes beyond pure radiative cooling, such as energy-dependent particle escape, changes in acceleration efficiency, or Klein--Nishina effects at the highest energies, likely contribute to shaping the electron spectrum \citep[e.g.,][]{2005MNRAS.363..954M, 2018MNRAS.478L.105J}.
The observed hardening of the broken power-law electron distribution indicates more efficient particle acceleration at higher flux levels, consistent with the well-known “harder-when-brighter” trend in blazars \citep[e.g.,][]{1998A&A...333..452K}.
In our modeling, the fitted break energy $\xi_b$ values lie in the range $(0.81$-$1.3)\times10^{-2}$, with the maximum value of $1.30\times10^{-2}$ obtained for the low-flux state. This shift implies more efficient radiative cooling during high-flux states than in the quiescent ones. The derived jet powers also support this picture, with a lower power of $3.09\times10^{45}\,\mathrm{erg\,s^{-1}}$ during the low-flux state and values up to $\sim10^{47}\,\mathrm{erg\,s^{-1}}$ in high-flux states. Furthermore, the model fits require relatively large bulk Lorentz factors, although their upper bounds remain poorly constrained across the different epochs. \

In summary, PKS~0402$-$362 exhibits pronounced multiwavelength variability and a double log-normal flux distribution, indicating the presence of distinct flux states in its $\gamma$-ray emission. A modulation on a timescale of $\sim$400 days is detected, but given the proximity of this timescale to one year and the limited number of observed cycles, we refer to this feature conservatively as a candidate for quasi-periodic modulation.
We emphasize that the reliable  results of this study arise from the broadband SED modeling. The systematic changes in magnetic field strength, particle spectra, and Compton dominance between flux states provide strong evidence for intrinsic variations in the jet physical conditions and are consistent with the harder-when-brighter behavior commonly observed in FSRQs. Continued high-cadence and long-term monitoring will be essential to further assess the nature of long-timescale variability in this source. Finally, although a one-zone leptonic model satisfactorily reproduces the time-averaged SEDs across all six states, the deviation of $\Delta p$ from the canonical cooling value,  motivates future time-dependent  modeling to fully disentangle acceleration, cooling, and escape processes.

%In summary, PKS\,0402$-$362 exhibits strong multiwavelength variability, a double log-normal flux distribution, and a significant $\sim$413-day quasi-periodic modulation in its $\gamma$-ray emission. The temporal, statistical, and spectral characteristics collectively indicate that the variability is governed by intrinsic jet dynamics modulated by geometric effects. Future high-cadence monitoring and long-term $\gamma$-ray observations will be essential to confirm the persistence of the detected periodicity and to constrain the physical mechanisms responsible for variability in powerful FSRQs.

%% Also note that the akcnowlodgment environment does not support long amounts of text. If you have a lot of people and institutions to acknowledge, do not use this command. Instead, create a new \section{Acknowledgments}.
\begin{acknowledgments}
Zeeshan Nazir is thankful to Department of Physics, Central University of Kashmir, Ganderbal 191201, India, for the support and facilities provided. SAD is thankful to the MOMA for the MANF fellowship (No.F.82-27/2019(SA-III)). ZS is supported by the Department of Science and Technology, Govt. of India, under the INSPIRE Faculty grant (DST/INSPIRE/04/2020/002319). ZM acknowledges the financial support provided by the Science and Engineering Research Board (SERB), Government of India, under the National Postdoctoral Fellowship (NPDF), Fellowship reference no. PDF/2023/002995. SAD,  ZS and ZM  express  gratitude to the Inter-University Centre for Astronomy and Astrophysics (IUCAA) in Pune, India, for the support and facilities provided.

\end{acknowledgments}


\vspace{5mm}


% \software{
%           }




\bibliography{sample631}{}
\bibliographystyle{aasjournal}



\end{document}